\documentclass[pra,twocolumn]{revtex4-1}
\usepackage{graphicx}
\usepackage{grffile}
\usepackage{dcolumn}
\usepackage{bm}
\usepackage{amsmath}
\usepackage{hyperref}
\hypersetup{colorlinks=true, urlcolor=blue}

\usepackage{hyperref}
\usepackage{epsfig}
\usepackage{newlfont}
\usepackage{amssymb}
\usepackage{amsfonts}
\usepackage{amsmath}
\usepackage{bm}
\usepackage{booktabs}

\def \be{\begin{equation}}
\def \ee{ \end{equation} }

\begin{document}
\renewcommand*{\DefineNamedColor}[4]{%
  \textcolor[named]{#2}{\rule{7mm}{7mm}}\quad
  \texttt{#2}\strut\\}

\definecolor{red}{rgb}{1,0,0}
\title{Local Decoherence-free Macroscopic Quantum states}
\author{Utkarsh Mishra, Aditi Sen(De), and Ujjwal Sen}

\affiliation{Harish-Chandra Research Institute, Chhatnag Road, Jhunsi, Allahabad 211 019, India}

\begin{abstract}
We identify a class of quantum states, each consisting of a microscopic and a macroscopic section, that are effectively decoherence-free 
when each particle is locally passed through 
a quantum channel. In particular, and in contrast to other macroscopic quantum states like the Greenberger-Horne-Zeilinger state, the content of entanglement and other quantum correlations in the microscopic to macroscopic partition of this class of states is independent of the number of particles in their macroscopic sectors, when all the particles suffer passage through local amplitude and phase damping channels. Decay of quantum correlations -- entanglement as well as quantum discord -- of this class of states in the microscopic to macroscopic partition is also much lower in the case of all the local quantum channels, as compared to the other macroscopic superposition states. The macroscopic sections of the states are formed, in each case, by using a Dicke state and an orthogonal product state, which are macroscopically 
distinct in terms of markedly different amounts of violation of Bell inequality. 
\end{abstract}
\maketitle
\section{Introduction}
Quantumness of a physical system plays an important role in several quantum communication and computational tasks \cite{NCHUANG}. Such protocols include quantum teleportation \cite{bennettelp}, super-dense
coding \cite{sudcoding}, quantum cryptography \cite{crypt}, and deterministic quantum computation with one qubit \cite{deterqc}. The advantages of these protocols over their classical counterparts vitally 
depend on the amount of quantum coherence present in the system. The physical system is almost always 
coupled to 
an environment, which, 
in general, destroys 
quantum coherence.  It is therefore crucially important to control and preserve the quantum character of physical systems in which quantum information protocols are to be implemented. 
A variety of experimental techniques have been developed in the last decade to isolate a quantum system from its environment in ion traps \cite {NRLEKB}, cold atoms \cite{JZ} and other systems \cite{SSEXPT}. 
Improvements in computational tasks by using quantum mechanics 
require
the creation of quantum coherence in  suitable macroscopic quantum 
systems which can reasonably withstand decoherence due to environmental noise.
Demonstration of macroscopic quantum 
states in various physical systems, like superconductors \cite{5R9N}, nanoscale magnets \cite{10F12W}, laser cooled trapped ions \cite{13M}, photons in a microwave cavity \cite{14B},  and \(C_{60}\)
molecules \cite{15A} have been proposed. 
Macroscopic quantum superposition in superconducting quantum interference devices (SQUID) \cite{FVCTL}  has already been experimentally achieved (see also \cite{othermacro}). It is important to identify systems which can 
retain their quantum coherence even with an increase in the  number of particles under decoherence.
Such study has importance also in fundamental questions like quantum to classical transition \cite{zurek}, existence of quantum superpositions at the  macroscopic level \cite{SCAT}, etc.

The effect of decoherence on  systems of many qubits have been studied by using different types of noise models. In particular, the effect of local depolarizing noise on the multiqubit  
Greenberger-Horne-Zeilinger (GHZ) state \cite{GHZ} has been investigated, and it has been found that the state remains entangled up to $55$\% noise for any number of qubits \cite{simonkempe}. 
(See \cite{multipartylnoise, utkarshpra} for further work in this direction.) It has also been shown that encoding of logical qubits into 
certain 
subspaces of larger number of physical qubits 
can lead to effectively decoherence-free qubits against certain types of noise models \cite{lidarwhaley}.

 In this paper, we identify a class of macroscopic quantum states which is robust against a large spectrum of physically reasonable local noise models --it
 is effectively decoherence-free for certain local noisy
 channels, and weathers decoherence better than other macroscopic states for the remaining channels. 
 The robustness of  entanglement \cite{eita-HHHH} and other quantum correlations \cite{eita-Kavan} 
of the macroscopic state is considered in a partition of the entire system into a macroscopic and a macroscopic part. 
To study decoherence, we consider three kinds of noise models: local phase damping, local amplitude 
 damping, and local depolarizing channels. We refer to the macroscopic states in the class as the \(H_{C_{N}}^{m}\) states, 
 which have $k$ particles in its microscopic part and $N$ in its macroscopic part. 
 The macroscopic sector is built by using $W$ states \cite{W1,W2,Gdansk}, or more generally, Dicke states \cite{dicke-state} with \(m\) 
 excitations, and by an orthogonal product state. 
%
The macroscopic parts of the states are macroscopically distinct in terms of their drastically different 
amounts of violation of Bell inequality \cite{Bell,SCAT,W2,utkarshpra}  (cf. \cite{multipartylnoise}). We show that for the \(H_{C_{N}}^{m}\) state, the entanglement as well as quantum discord in the bipartition into microscopic and macroscopic parts 
do not depend on the number of particles in the macroscopic sector, when all the particles are sent individually through local phase damping or local amplitude damping channels. 
The same behavior is not true for the local depolarizing noise,  although the  \(H_{C_{N}}^{m}\) state withstands the noise better than other macroscopic states. 

We also find that the \(H_{C_{N}}^{m}\) state can sustain its quantum correlations better than other macroscopic superposition states
against the local noisy channels. In particular, the \(H_{C_{N}}^{m}\) state, e.g. for \(m=1\) and \(N=6\), can
withstand approximately $97\%$ noise in both cases, until when its entanglement in the micro : macro bipartition 
  remains non-vanishing, while the  values are \(78\%\) and \(67.5\%\) respectively for the corresponding GHZ state. 
  We also compare the effects of the local noise models on the \(H_{C_{N}}^{m}\) state with those on other macroscopic states which have 
  unit entanglement in the micro : macro bipartition, just like the \(H_{C_{N}}^{m}\) state.
  Such investigations can shed light on our quest towards identifying 
  appropriate memory devices involving a large number of qubits.

The paper is structured as follows. In Sec. \ref{noise-model}, we introduce  the local noise models,
while the measures of quantum correlation are defined in Sec. \ref{q-c}. The  
class of macroscopic states is introduced in Sec. \ref{macro-state-dec}. Their quantum correlation properties under the 
local phase damping channel (LPDC), the local amplitude damping channel (LADC),
and the local depolarizing channel (LDPC) are discussed in Secs.  \ref{macro-state-dec-pdc},  \ref{macro-state-dec-adc}, and  \ref{macro-state-dec-dpc} respectively. 
The effects of the same local channels on the GHZ state is reported in Sec. \ref{ghz-decoh}, where a comparison of the 
\(H_{C_{N}}^{m}\) state with the GHZ state is also considered.
We compare 
the effects of the local noises on further macroscopic states with the new class of states in Sec. \ref{other-macro-states}. We conclude in Sec. \ref{conclusion}.

\section{Quantum channels} 
\label{noise-model}

In this section, we will discuss about the local decoherence models. In particular, we consider three types of noisy environments, or equivalently consider that the system passes through the corresponding quantum channels \cite{preskill}, viz.,
\((a)\)  phase damping, 
\((b)\)  amplitude damping,  and
\((c)\)  depolarizing channels.
The entire system will be considered to be a collection of qubits, in some initial state, and we will consider the effect after each of the qubits pass through one of the quantum channels.

\subsection{Phase damping channel}
Phase damping happens when e.g., a photon travels through a waveguide, and scatters randomly. An initial single qubit state \(\rho\) evolves under phase damping as
\begin{equation}
 \rho\mapsto\rho'=(1-p)\rho+M_{1}\rho M_{1}+M_{2}\rho M_{2}.
\end{equation}
where 
\(M_{1}=\sqrt{p}\arrowvert 0\rangle\langle 0 \arrowvert\) and \(M_{2}=\sqrt{p}\arrowvert 1\rangle\langle 1 \arrowvert\) with \(p\) being a probability. Hence under this noise model, the final state becomes 
\[\rho'= \left( \begin{array}{cc}
\rho_{00} & (1-p)\rho_{01} \\
(1-p)\rho_{10}&\rho_{11} \end{array} \right),\]
where \(\rho_{ij},i,j=0,1\), are the matrix elements of the initial state \(\rho\). Here $\arrowvert 0\rangle$ and $\arrowvert 1 \rangle$ are eigenstates of the Pauli matrix $\sigma_{z}$,
with eigenvalues $-1$ and $1$ respectively.
 The diagonal terms of the initial density matrix remain invariant under the phase damping channel while the off-diagonal terms decay with probability \((1-p)\). 

\subsection{Amplitude damping channel}
The amplitude damping channel is a model for the decay of an excited state of a (two-level) atom due to spontaneous emission of photons.
Detection of the emitted photon (``observation of the environment''), via a positive operator valued measurement, gives us information about the initial preparation of
the atom.
After passing through an amplitude damping channel, the initial qubit state \(\rho\) is transformed to 
\[\rho'= \left( \begin{array}{ccc}
\rho_{00}+p\rho_{11} & \sqrt{(1-p)}\rho_{01} \\
\sqrt{(1-p)}\rho_{01}&(1-p)\rho_{11} \end{array} \right).\]
where \(p\) is the rate of decay. Unlike the phase damping channel, both diagonal and off-diagonal terms gets affected by this channel.
\subsection{Depolarizing channel}

The errors which happen to an arbitrary  pure qubit, say \(\arrowvert \Psi\rangle\), when interacting with its environment, 
can be categorized into
bit flip error, which transforms  \(\arrowvert\Psi\rangle\) into \(\sigma_x \arrowvert\Psi\rangle\), 
 phase flip error, which transforms  \(\arrowvert\Psi\rangle\) into \(\sigma_z \arrowvert\Psi\rangle\), and
bit-and-phase-flip error. Here \(\sigma_x, \sigma_y, \sigma_z \) are the three Pauli matrices.
If an arbitrary qubit \(\rho\) is sent through a depolarizing channel, the state remains unchanged with probability \((1-p')\) while the above three kinds of error occur 
with probability \(\frac{p'}{3}\) each. Therefore, \(\rho\), sent through a depolarizing channel, transforms as
\begin{equation}
\label{eq:dpc}
\rho\mapsto\rho'=(1-p')\rho+\frac{p'}{3}(\sigma_{x}\rho\sigma_{x}+\sigma_{y}\rho\sigma_{y}+\sigma_{z}\rho\sigma_{z}).
\end{equation}
By putting \(p'\)=\(\frac{3p}{4}\), we obtain
\begin{equation}
\label{eq:white-noise}
 \arrowvert i\rangle\langle j \arrowvert\mapsto\frac{p}{2} I \mbox{tr}(\arrowvert i\rangle\langle j \arrowvert)+(1-p)\arrowvert i\rangle\langle j \arrowvert,
\end{equation}
where \(I\) is the identity operator on the qubit Hilbert space. Here, $0\leq p'\leq \frac{3}{4}$ and $0\leq p \leq 1$. 
Note that for \(p'\)=\(\frac{3}{4}\), the qubit state in Eq. (\ref{eq:dpc}) will be proportional to the identity matrix, \(I\), 
while the same for the noisy state obtained via  Eq. (\ref{eq:white-noise}) occurs at  $p=1$.

\section{Quantum correlations}
\label{q-c}
It is important to understand the nature and content of quantum correlations that a superposed  state of a composite system of a large number of particles can retain, when the constituent particles are sent through noisy channels.  
For such an investigation, we define two quantum correlation measures -- logarithmic negativity \cite{vidal01}, 
for quantifying entanglement, and quantum discord \cite{discord}, a quantum correlation measure that is independent of the entanglement-separability paradigm.

\subsection{Logarithmic negativity}
\label{log-neg}

A computable measure of entanglement in a bipartite state, \(\rho_{AB}\), shared between \(A\) and \(B\), is the 
logarithmic negativity 
\cite{vidal01}, defined as 
\begin{equation}
\label{eq:def-neg-ln}
 E_{N}(\rho) = \log_2(2N(\rho)+1),
\end{equation}
where the ``negativity'', \(N(\rho)\), is 
the sum of the absolute values of the negative eigenvalues of the partially transposed 
state, \(\rho^{T_{A}}_{AB}\) where the partial transposition is taken with respect to \(A\) \cite{Peres-Horodecki}. 

\subsection{Quantum discord}
Quantum discord is a measure of quantum correlation based on information-theoretic concepts and is independent of entanglement. Such quantification is at least partly induced by the discovery, over the last decade, of  several 
non-classical phenomena which can not be explained by using entanglement \cite{discordappl1,deterqc,discordappl3}.

There are two equivalent ways to define the mutual information between two classical random variables. 
These two classically equivalent definitions of mutual information, after ``quantization'',  
produce two inequivalent concepts, the difference of which is  termed as the quantum discord \cite{discord}. 

Quantization of one of the classical definitions leads to the ``quantum mutual information'', which, for a bipartite 
quantum state, \(\rho_{AB}\),
is defined as \cite{qmi} (see also \cite{Cerf, GROIS})
\begin{equation}
\label{qmi1}
I(\rho_{AB})= S(\rho_A)+ S(\rho_B)- S(\rho_{AB}),
\end{equation}
with \(\rho_A\) and \(\rho_B\) being the local density matrices of \(\rho_{AB}\), and 
where \(S(\sigma) = - \mbox{tr} \left(\sigma \log_2 \sigma\right)\) is the von Neumann entropy of the quantum state \(\sigma\).

Quantization of the other  definition of classical mutual information leads to the quantity 
\begin{equation}
\label{cmi1}
 J(\rho_{AB}) = S(\rho_A) - S(\rho_{A|B}),
\end{equation}
where the ``quantum conditional entropy'', \(S(\rho_{A|B})\equiv S_{A|B}\), is defined as  
\begin{equation}
\label{qce}
S(\rho_{A|B}) = \min_{\{\Pi_i^B\}} \sum_i p_i S(\rho_{A|i}),
 \end{equation}
where the minimization is performed  over all 
rank-1
measurements, \(\{\Pi^B_i\}\),  performed on subsystem \(B\).
Here, \(p_i = \mbox{tr}_{AB}(\mathbb{I}_A \otimes \Pi^B_i \rho_{AB} \mathbb{I}_A \otimes \Pi^B_i)\) is the probability for obtaining the outcome \(i\), and 
the corresponding post-measurement state 
for the subsystem \(A\) is \(\rho_{A|i} = \frac{1}{p_i} \mbox{tr}_B(\mathbb{I}_A \otimes \Pi^B_i \rho_{AB} \mathbb{I}_A \otimes \Pi^B_i)\), 
where \(\mathbb{I}_A\) is the identity operator on the Hilbert space of the quantum system that is with \(A\). The quantity 
\( J(\rho_{AB})\) has been  argued to be the amount of  classical correlations in  \(\rho_{AB}.\)

In this paper, we calculate the logarithmic negativity and the quantum discord of the decohered macroscopic states in the micro : macro bipartition. For the case of quantum discord, the measurements (to evaluate \(J\)) are 
carried out in the micro part.

\section{Macroscopic State under Decoherence}
\label{macro-state-dec}
In this  section, we first introduce a class of macroscopic quantum states. We then address two aspects of these states:
(1) we consider the effects of local decoherence on  these states and identify 
the state which is more robust against local noise than the other states, and (2) we study the scaling behavior of quantum correlations of this class of states with the increase in number of particles against noise.
\subsection{A class of macroscopic quantum states}
\label{macro-state-def}
Let us  define here a class of quantum states, each consisting of a microscopic and a macroscopic part. We denote it by \(\arrowvert H_{C_{N}}^{m}\rangle\), and it is given by 
\begin{equation}
\label{eq:hcs-class-state}
 \arrowvert H_{C_{N}}^{m}\rangle=\frac{1}{\sqrt{2}}[\arrowvert 0^{\otimes k}\rangle_{\mu}\arrowvert W_{N}^{m}\rangle_{M}+\arrowvert 1^{\otimes k}\rangle_{\mu} \arrowvert 0^{\otimes N}\rangle_{M}],
\end{equation}
where
\begin{equation}
\label{eq:wnm}
 \arrowvert W_{N}^{m}\rangle=\frac{1}{\sqrt{\binom {N} {m}}}\sum\arrowvert 1^{\otimes m}0^{\otimes N-m}\rangle.
\end{equation}
The sum in the last equation denotes the equal superposition of all the $\binom {N} {m}$  combinations of \(m\) \(\arrowvert 1\rangle\)'s and (\(N-m\)) \(\arrowvert 0\rangle\)'s. Here $\binom {N} {m}=\frac{N!}{m!(N-m)!}$. 
The suffix \(\mu\) denotes the microscopic part while the suffix \(M\) is for  the macroscopic sector of the state. We assume that $1\leq m <N$. 
The case $m=0$ is uninteresting,
as then the $\mu$ : $M$ partition is unentangled. The  \(H_{C_{N}}^{m}\) state becomes a GHZ state for $m=N$,  which is considered separately in the succeeding section. We will generally be interested in the cases where  \(1\leq k\ll N\), i.e., where 
the number of particles (qubits) in 
the microscopic part is much smaller than that of the macroscopic part. For $k=1$ and $m=1$, this reduces to the $H_{C}$ state \cite{rafelchaves,utkarshpra}.

For investigating the quantum coherence of this class of quantum states, each qubit of the  \(H_{C_{N}}^{m}\) state is sent through a noisy quantum channel. We then investigate the 
behavior of entanglement and quantum discord in the microscopic to macroscopic bipartition.

The state $\arrowvert H_{C_{N}}^{m}\rangle$  has unit entanglement in the $\mu$ : $M$ 
partition, and is of the form of the  Schr\"{o}dinger cat state, i.e., $\arrowvert \bar{0}\rangle_{\mu}\arrowvert \mbox{alive}\rangle_{M}+\arrowvert \bar{1}\rangle_{\mu}\arrowvert \mbox{dead}\rangle_{M}$, 
where $\arrowvert \bar{0}\rangle_{\mu}$ and $\arrowvert \bar{1}\rangle_{M}$ 
are orthonormal states of the microscopic part, and $\arrowvert \mbox{alive}\rangle_{M}$ and 
$\arrowvert \mbox{dead}\rangle_{M}$ are orthonormal states of the macroscopic (cat) part. In the case of 
\(H_{C_{N}}^{m}\), the ``alive" and ``dead" parts are respectively modeled by $\arrowvert W_{N}^{m}\rangle$ and $\arrowvert 0^{\otimes N}\rangle$, with the latter being macroscopically 
distinct in terms of their violation of Bell inequalities \cite{W2,utkarshpra}.

\subsection{ \(H_{C_{N}}^{m}\) state under local phase damping channel}
\label{macro-state-dec-pdc}
Let us begin with the situation when  each qubit of the  \(H_{C_{N}}^{m}\) state is sent through a  phase damping channel. 
The block of the local phase damped \(H_{C_{N}}^{m}\) state, after partial transposition, which contributes
in the calculation of
logarithmic negativity, 
in the micro : macro bipartition, is of the form
\[B_{H_{C_{N}}^{m}}^{lpdc}= \frac{1}{2}\left( \begin{array}{cccccc}
0 & b&.&.&.&b \\
b&0&.&.&.&0 \\
.&.&.&.&.&.\\
.&.&.&.&.&.\\
.&.&.&.&.&.\\
b&0&.&.&.&0
\end{array} \right),\]
where \(b=\frac{1}{\sqrt{\binom{N}{m}}}(1-p)^{k+m}\).
This block has one negative eigenvalue, given by
\begin{equation}
\label{eq:neg-pd-hcs} 
 \lambda_{H_{C_{N}}^{m}}^{lpdc}=-\frac{1}{2}(1-p)^{k+m}.
\end{equation}
Therefore, the logarithmic negativity of the local phase damped state is  given by
\begin{align}
\label{eq:lpdcnm}
 E_{H_{C_{N}}^{m}}^{lpdc}(k,m)=\log_{2}[2|\lambda_{\rho_{H_{C_{N}}^{m}}}^{lpdc}|+1].
\end{align}

\begin{figure}
 \includegraphics[angle=270,width=4.2cm,totalheight=3.7cm]{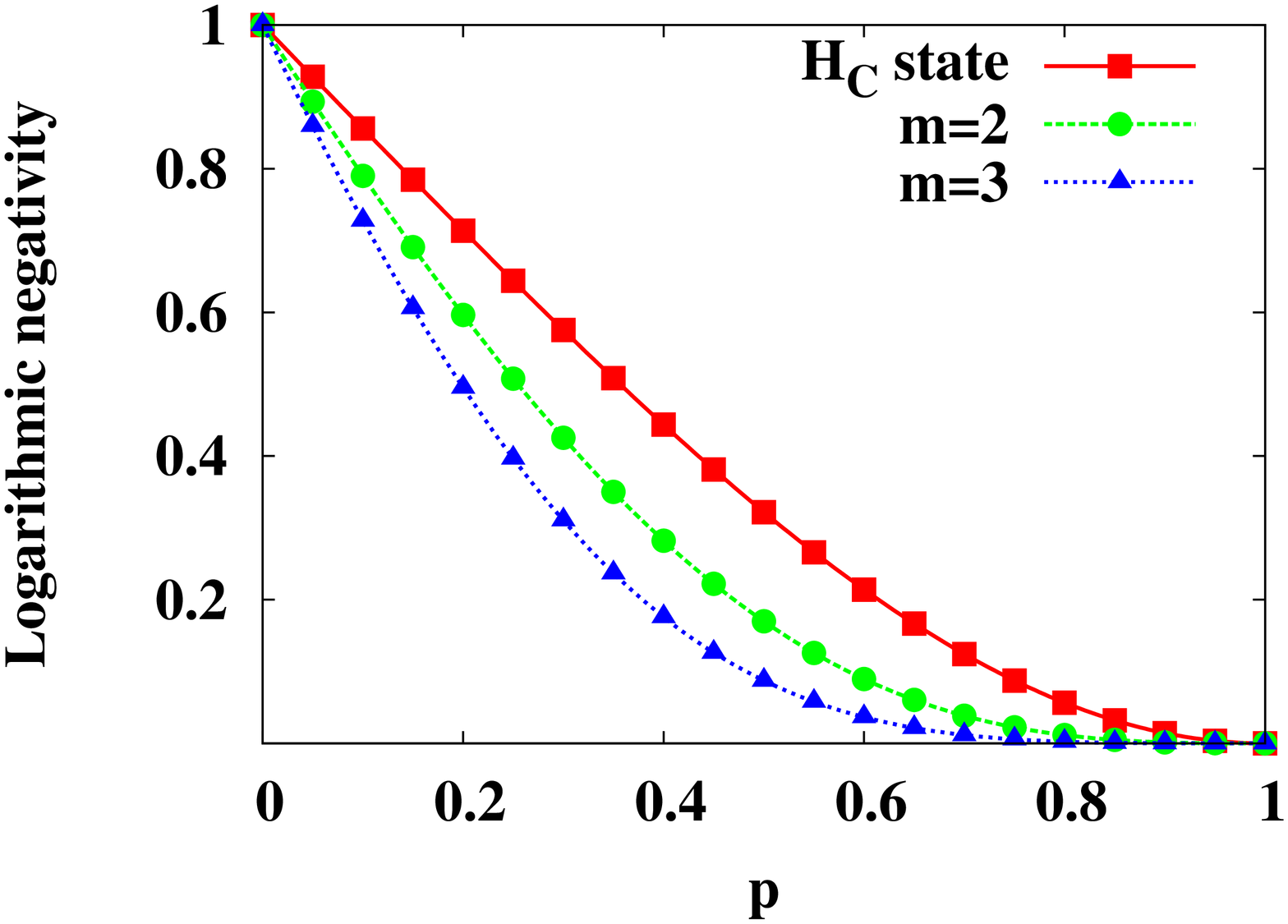}
  \includegraphics[angle=270,width=4.2cm,totalheight=3.7cm]{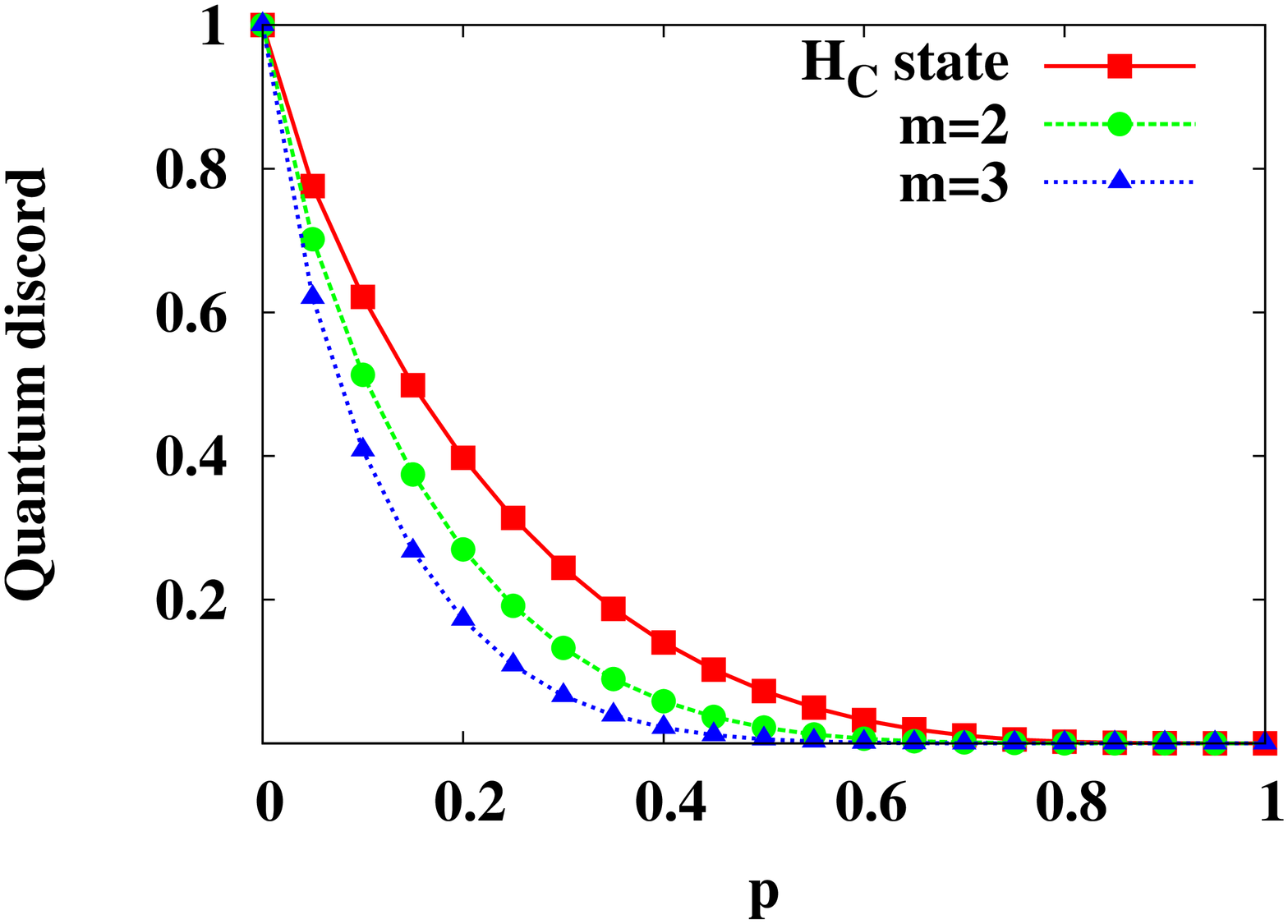}
 \caption{(Color online) 
 Entanglement and quantum discord of the local phase damped \(H_{C_{N}}^{m}\) state. 
 We  plot the logarithmic negativity (in ebits) and quantum discord (in bits) on the vertical axes versus the decoherence parameter \(p\) (dimensionless)
 on the horizontal axes, for the \(H_{C_{N}}^{m}\) state, after each of the qubits are affected by phase damping noise. 
 The plot are displayed for \(m=1\), 2, and 3. Note that for \(m=1\), the \(H_{C_{N}}^{m}\) state is the same as the \(H_C\) state. All plots are 
 for \(N=6\) and for \(k=1\). 
 Here, and in the rest of the paper, we mostly plot the curves for the different quantities for a modest 
 number of particles in the macroscopic 
 sector (which, in the current case implies that we are dealing with a \(2^7 \times 2^7\) matrix). This is despite the fact that in many cases, 
 we can consider bigger system sizes and even have analytical results for arbitrary \(N\). We however feel that the curves for the relatively modest 
 system sizes will give the reader a feeling of the situation in a potential experimental realization of the phenomena considered.
 }
 \label{fig:EXACT_LN_HCS_PDC}
\end{figure}

An important point to note is that the entanglement does not depend on the total number of particles, \(N\). That is, the effect on the \(H_{C_{N}}^{m}\) state after all $N$ qubits of the 
state are sent through phase damping channels, is independent of the size of the macroscopic part. As we will see below, (Sec. \ref{macro-state-dec-adc}), this beautifully simple situation persists for the 
local amplitude damping channel. The case is richer for the local depolarization (Sec. \ref{macro-state-dec-dpc}), and the noise-affected state does depend on $N$, although the scaling of quantum 
correlations is better than in the noise-affected GHZ state. Coming back to local phase damping, when $m=1$, $k=1$, we obtain
\begin{equation}
\label{eq:ln-pd-hcs} 
 E_{H_{C_{N}}^{m}}^{lpdc}(1,1)=\log_{2}(\gamma^{2}+1),
\end{equation}
with \(\gamma=1-p\). This  value of entanglement is the maximum among all other states in this class, i.e., among all \(H_{C_{N}}^{m}\), as also observed in 
Fig. \ref{fig:EXACT_LN_HCS_PDC}. Moreover, note that in the noiseless case, the entanglement is unity for all  \(H_{C_{N}}^{m}\).
\begin{figure}
 \includegraphics[angle=270,width=8.2cm,totalheight=4.7cm]{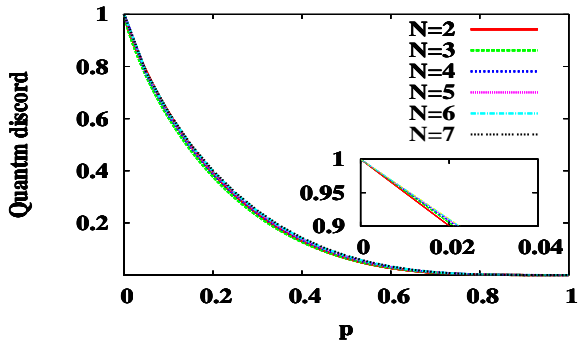} 
 \caption{(Color online) Independence of quantum discord of the local phase damped \(H_{C_{N}}^{1}\) state on the number of 
 particles in the macroscopic sector. Quantum discord (in bits) for the \(H_{C_{N}}^{1}\) state (i.e., the \(H_C\) state) for 
 different values of \(N\) after the state is local phase damped, is plotted on the vertical axis, against the decoherence parameter \(p\)
 (dimensionless) on the horizontal axis. 
Here, the  microscopic part of the state consists
 of a single particle (i.e., \(k=1\)). This shows that the quantum discord, like entanglement, is independent, 
 up to numerical accuracy, of size of the macroscopic sector.
 A similar feature holds for higher values of \(k\). The inset shows the same figure blown up near \(p=0\). 
 }
 \label{fig:qd-pdc-diff-N}
 \end{figure}

Instead of entanglement, if one considers quantum discord, in the micro : macro bipartition, we again find 
that the \(H_{C_{N}}^{1}\) state, with a single particle in the microscopic section, is
maximally robust under this kind of noise than all the other states in this class (see Fig. \ref{fig:EXACT_LN_HCS_PDC}). Moreover, numerical simulation 
indicates
that quantum discord for this class of states is also independent of system-size of macroscopic part up to first order of 
magnitude  (see Fig. \ref{fig:qd-pdc-diff-N}). It is also observed that the independency of quantum discord on the number of parties in the macroscopic sector remains valid for higher values of \(k\).


\subsection{\(H_{C_{N}}^{m}\) state under local amplitude damping channel}
\label{macro-state-dec-adc}
\label{macro}
\begin{figure}
 \includegraphics[angle=270,width=8.2cm,totalheight=4.7cm]{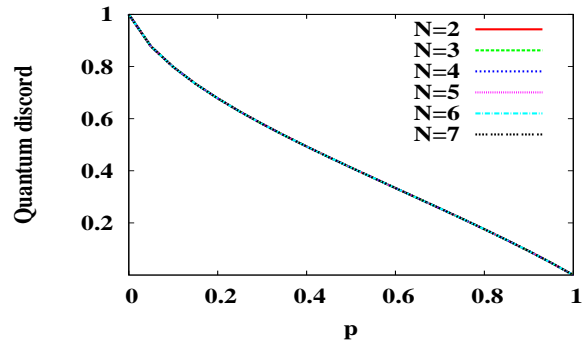}
 \caption{(Color online) Independence of quantum discord of the local amplitude damped \(H_{C_{N}}^{m}\) state 
 on the number of particles in the macroscopic sector. 
 All other considerations are the same as in Fig. \ref{fig:qd-pdc-diff-N}, except that we do not have the inset here.
%
}
 \label{fig:qd-adc-diff-N}
 \end{figure}

\begin{figure}
 \includegraphics[angle=270,width=4.2cm,totalheight=3.7cm]{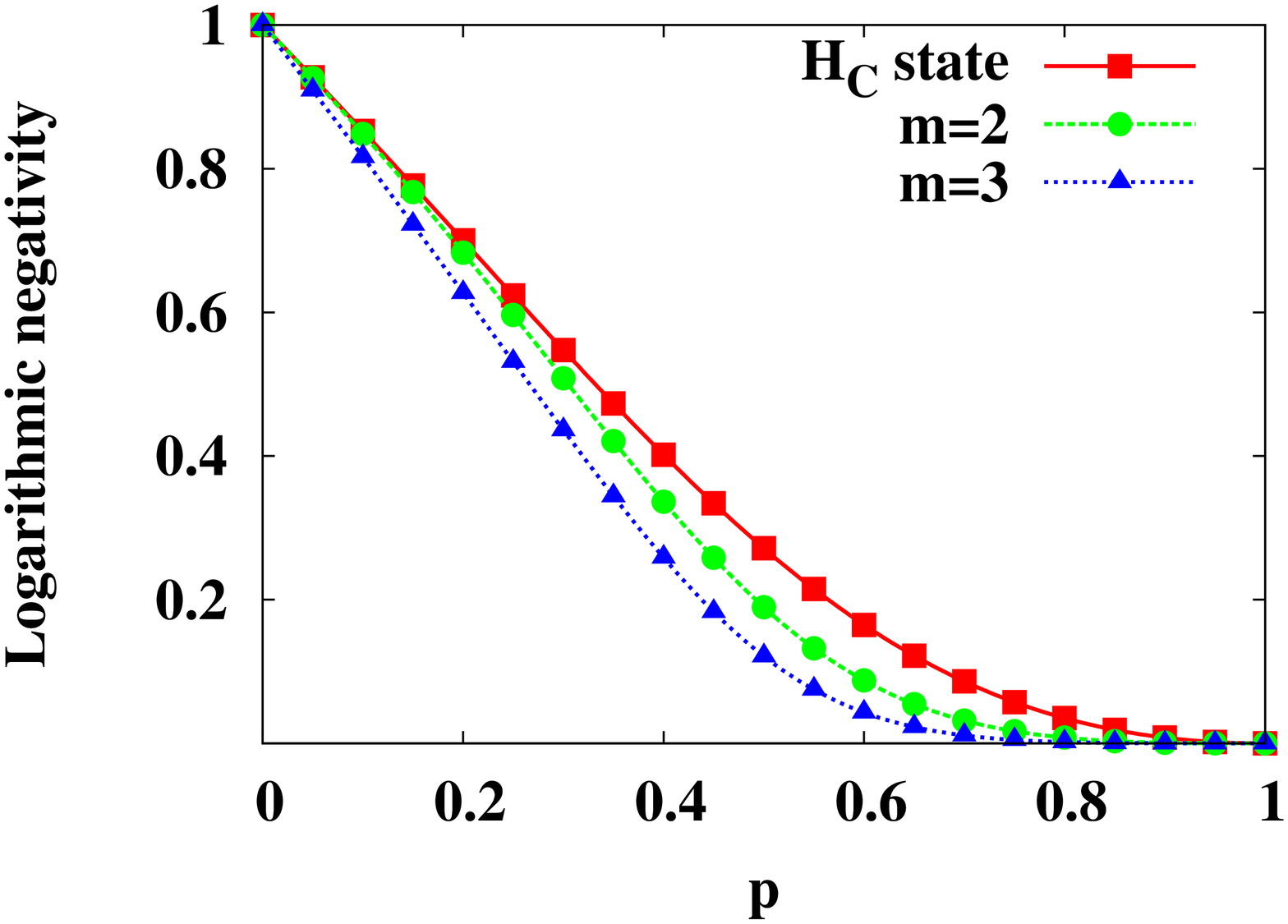}
 \includegraphics[angle=270,width=4.2cm,totalheight=3.7cm]{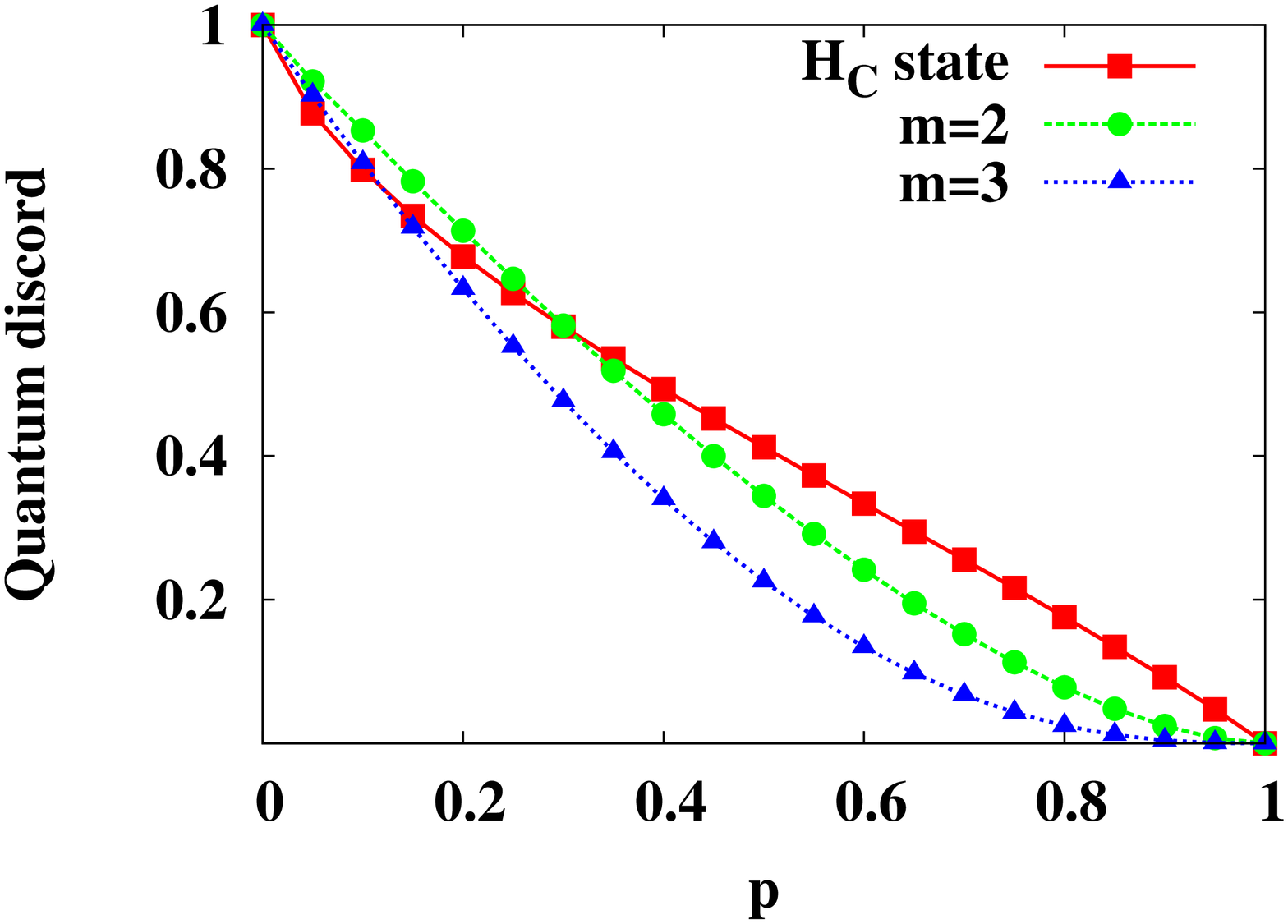}
 \caption{(Color online) Logarithmic negativity and quantum discord against the noise parameter of amplitude damping channel, for the 
 local amplitude damped \(H_{C_{N}}^{m}\) state. 
 All other considerations are the same as in Fig. 
 \ref{fig:EXACT_LN_HCS_PDC}.}
  \label{fig:ln-adc-hcs}
 \end{figure}

Consider now the situation when all the qubits of the  \(H_{C_{N}}^{m}\) state are sent through amplitude damping channels. In this case, the block of the partially transposed 
local amplitude damped \(H_{C_{N}}^{m}\) state, which gives negative eigenvalues contributing to  entanglement in the  micro : macro bipartition is an \((\binom {N} {m}+1)\times(\binom {N} {m}+1)\) matrix, and is given by
\[B_{H_{C_{N}}^{m}}^{ladc}=\frac{1}{2} \left( \begin{array}{cccccc}
p^{k}+p^{m} & \frac{(1-p)^{\frac{m+k}{2}}}{\sqrt{\bar{N}}}&.&.&.&\frac{(1-p)^{\frac{m+k}{2}}}{\sqrt{\bar{N}}} \\
\frac{(1-p)^{\frac{m+k}{2}}}{\sqrt{\bar{N}}}&0&.&.&.&0 \\
.&.&.&.&.&.\\
.&.&.&.&.&.\\
.&.&.&.&.&.\\
\frac{(1-p)^{\frac{m+k}{2}}}{\sqrt{\bar{N}}}&0&.&.&.&0
\end{array} \right).\]
Here, \(\bar{N}=\binom {N} {m}\). The negative  eigenvalue of this matrix, denoted by 
\(\lambda_{H_{C_{N}}^{m}}^{ladc}\),
is given by
\begin{align}
\label{eq:hcs-ampdc}
 \lambda_{H_{C_{N}}^{m}}^{ladc}=\frac{1}{4}( p^{k}+p^{m}-\sqrt{(p^{k}+p^{m})^{2}+4(1-p)^{k+m}} ),
\end{align}
and therefore the logarithmic negativity is 
\begin{equation}
\label{eq:ln-eq-adc-hcs}
 E_{H_{C_{N}}^{m}}^{ladc}(k,m)=\log_{2}[2|\mbox{min}(0,\lambda_{H_{C_{N}}^{m}}^{ladc})|+1].
\end{equation} 
It is clear from Eqs. (\ref{eq:hcs-ampdc}) and (\ref{eq:ln-eq-adc-hcs}) that the logarithmic negativity is independent of \(N\), just as for the local phase damping channel. 
Quantum discord, which is obtained numerically, is  also independent of system size, as depicted in Fig. \ref{fig:qd-adc-diff-N}, 
for \(k=1\). The independency of quantum discord on \(N\) holds true also  for higher values of \(k\). 
For \(k=1\) and \(m=1\), the  negative eigenvalue in Eq. (\ref{eq:hcs-ampdc}) reduces to 
\begin{equation}
\label{eq:hcnm1adc}
 \lambda_{H_{C_{N}}^{m}}^{ladc}(1,1)=\frac{1}{4}(2p^{2}-\sqrt{4p^{2}+4(1-p)^{2}}).
\end{equation}
Just like for the local phase-damping channel, the state \(H_{C_{N}}^{1}\) can sustain more noise than any other states in this class for both the quantum correlation measures. Note here that quantum discord is non-zero  
for the  entire range of the noise parameter, except at \(p=1\). 
Both the quantum correlations are plotted in Fig. \ref{fig:ln-adc-hcs} for different values of \(m\). 

 \subsection{\(H_{C_{N}}^{m}\) state under local depolarizing channel}
  \label{macro-state-dec-dpc}
We now consider the effect of the local depolarizing channel on the \(H_{C_{N}}^{m}\) state. Unlike phase and amplitude damping channels, entanglement in this case does depend on
the total number of particles, \(N\), in the macroscopic part, and decreases with the increase of \(N\).
We also probe the behavior of entanglement with respect to \(m\) for fixed total number of particles and also for a fixed number, $k$, of parties in the microscopic part. 
\begin{figure}
 \includegraphics[angle=270,width=4.2cm,totalheight=3.7cm]{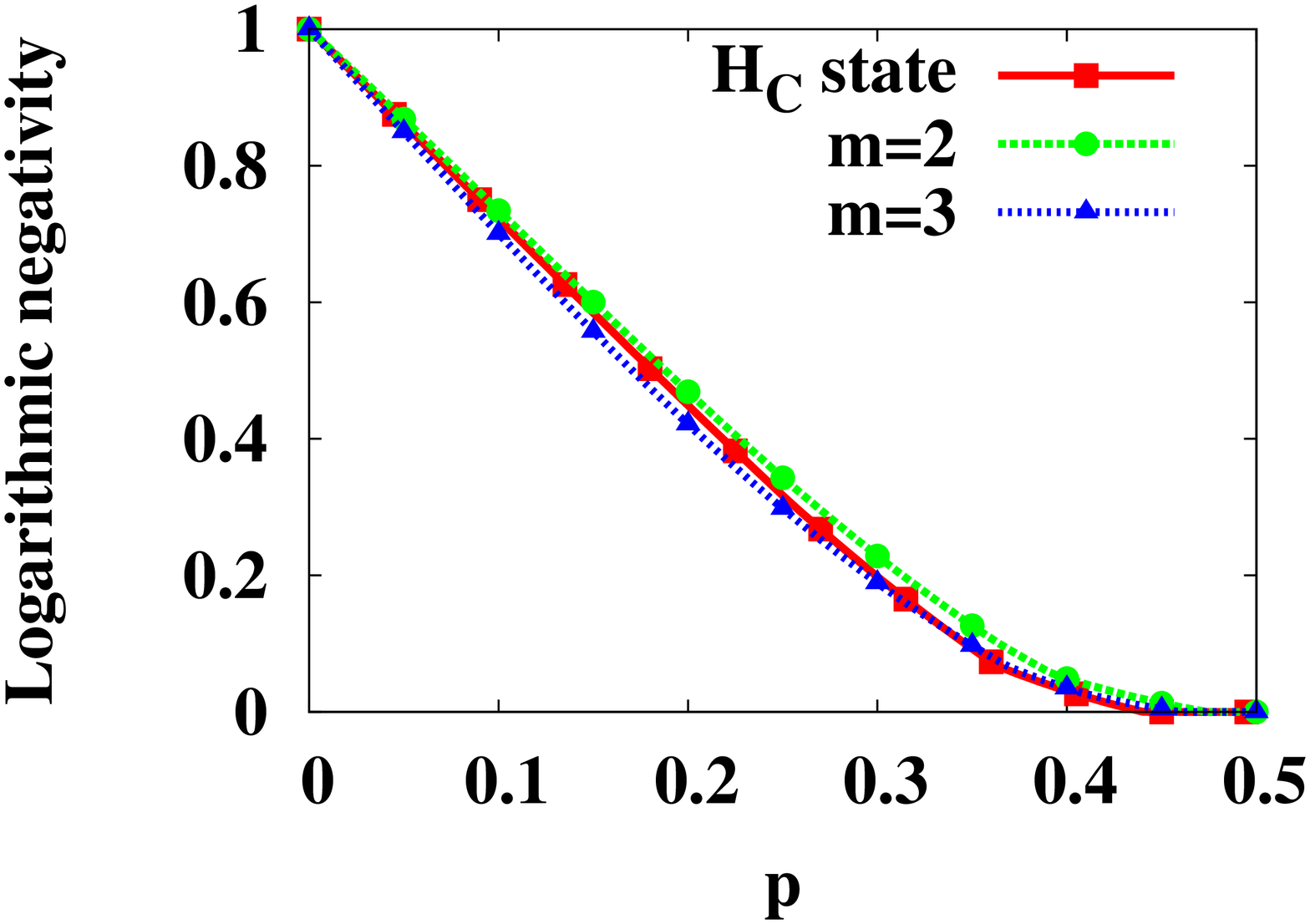}
  \includegraphics[angle=270,width=4.2cm,totalheight=3.7cm]{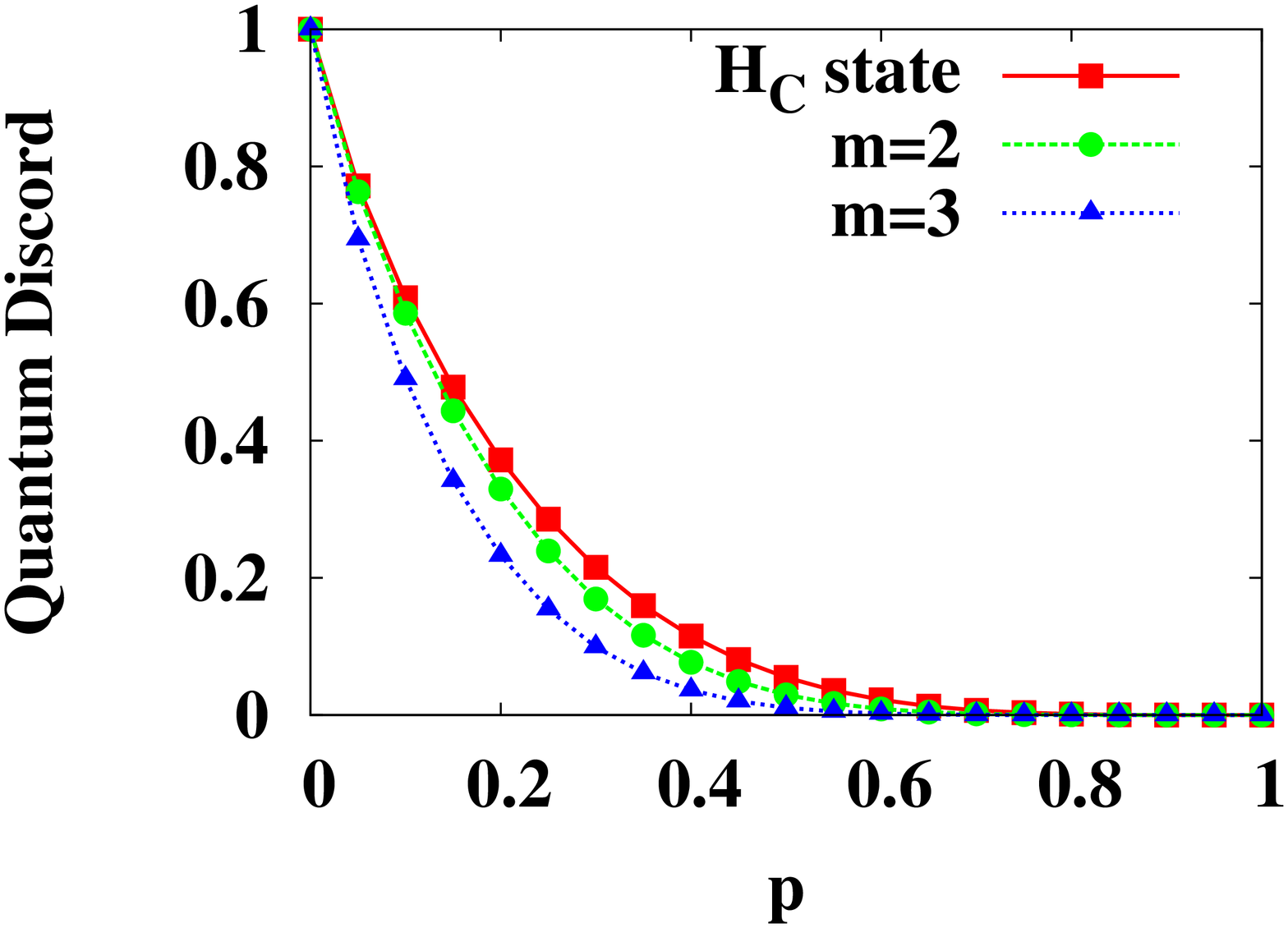}
 \caption{(Color online) Logarithmic negativity  and quantum discord for the local depolarized \(H_{C_{N}}^{m}\) state 
 against the noise parameter of the depolarizing channel. All other considerations are the same as in Fig.  \ref{fig:EXACT_LN_HCS_PDC}.
}
 \label{fig:ln-dpc-hcs}
 \end{figure}
 As seen in Fig. \ref{fig:ln-dpc-hcs} (left), the state with two  excitations (i.e., $m=2$), is more robust against local depolarizing channels than the state with one 
  excitation (i.e., $m=1$). Likewise, quantum discord decreases with the increase of excitations  (see Fig. \ref {fig:ln-dpc-hcs} (right)).

\subsection{A comparison of the local decohering channels}
 
\begin{figure}
 \includegraphics[angle=270,width=4.2cm,totalheight=3.7cm]{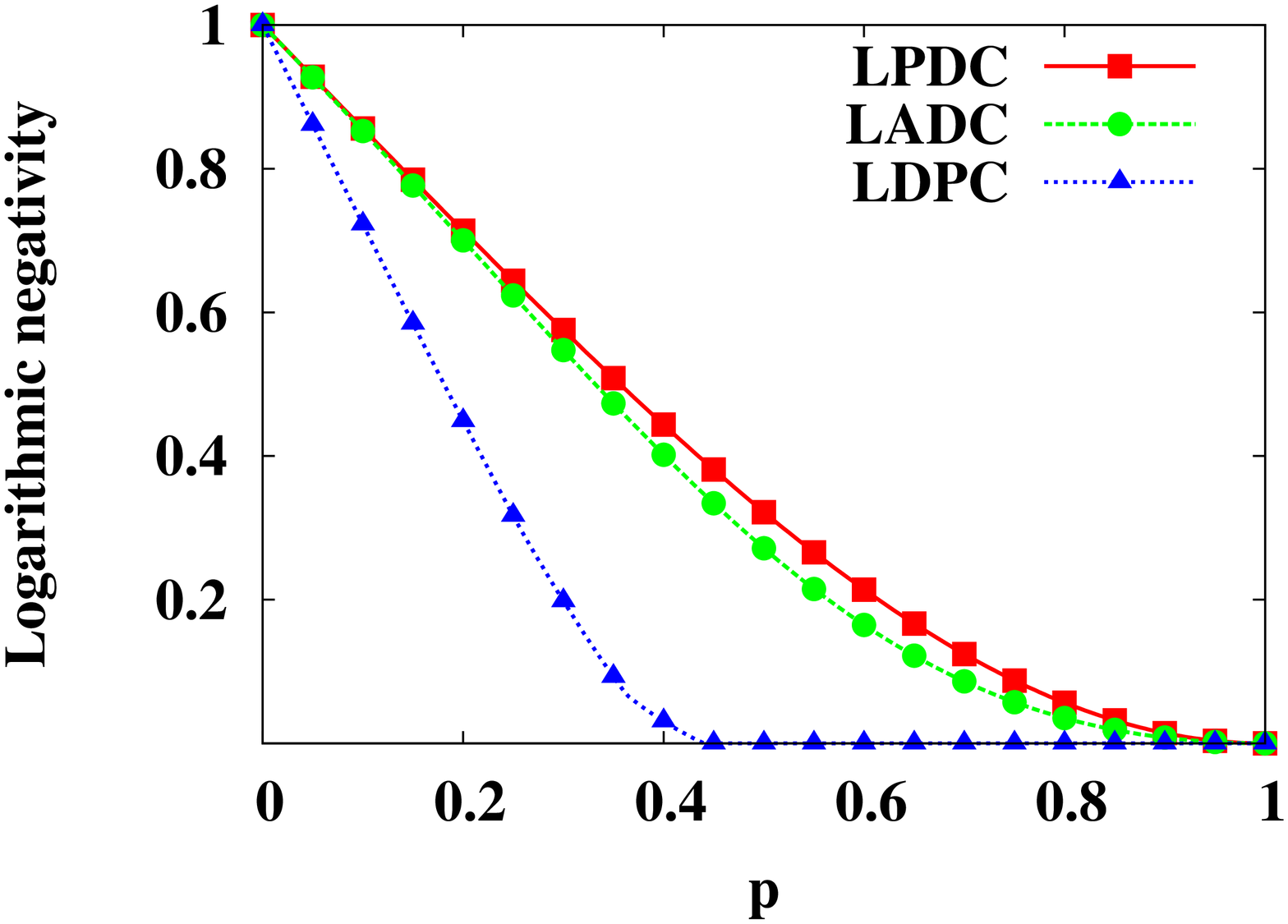}
  \includegraphics[angle=270,width=4.2cm,totalheight=3.7cm]{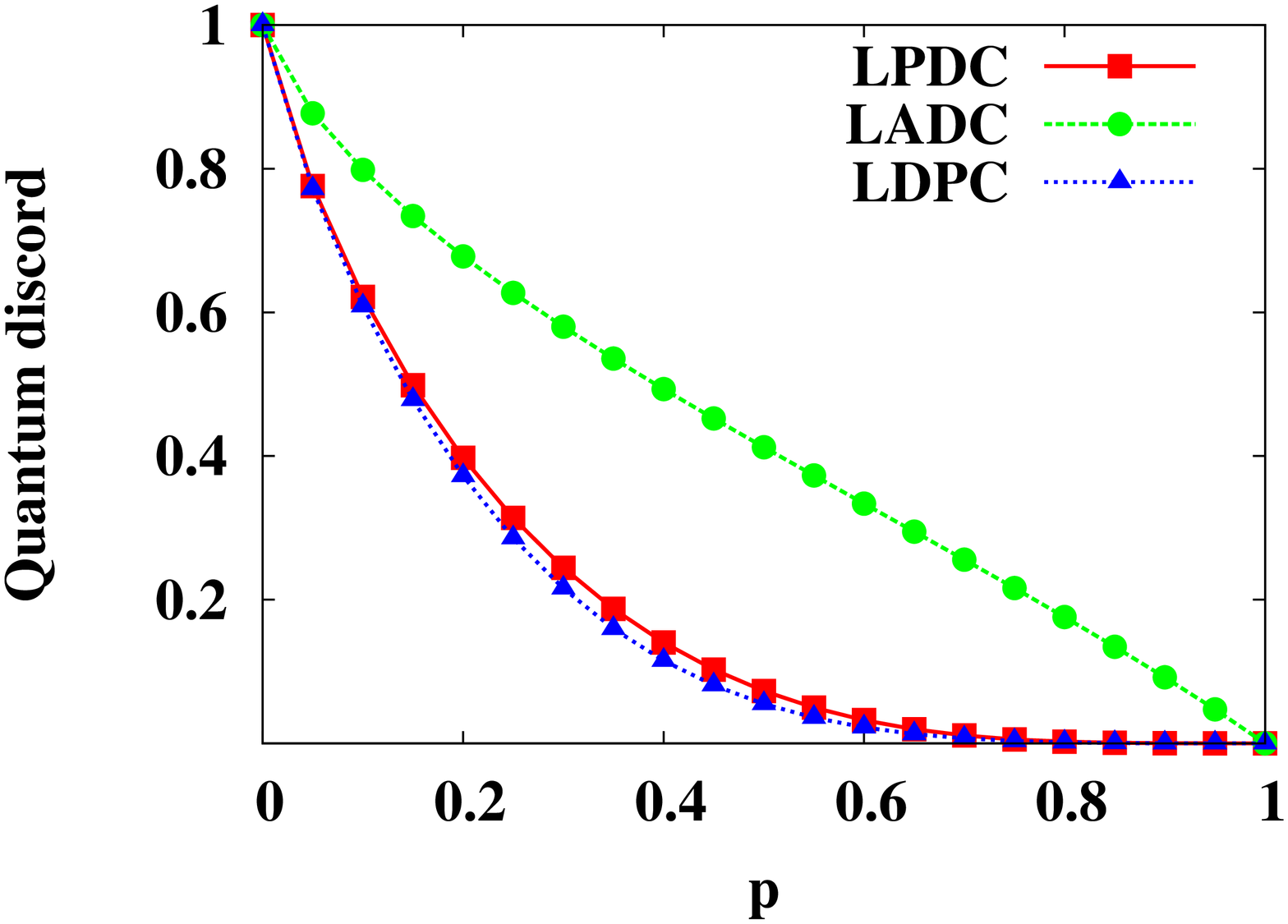}
 \caption{(Color online) How much quantum correlation is retained against which noise? 
 Logarithmic negativity and quantum discord for different noisy channels in the micro : macro bipartition for the \(H_{C_{N}}^{1}\) state for \(k=1\). Here we have taken \(N=6\). The vertical axes are in ebits for logarithmic negativity and in bits for quantum discord.
 The dimensionless parameter \(p\) on the horizontal axes corresponds to the noise parameter in the LPDC, the LADC, or the LDPC.}
 \label{fig:ln-pdc-adc-dpc-hcs}
 \end{figure}

 It is interesting to find the channel, from the ones considered in this paper, that is least destructive for the \(H_{C_{N}}^{m}\).
 For fixed \(N\), \(m\), and $k$, we compared their effects on entanglement and discord of the state. An example of such comparison is presented in Fig. \ref{fig:ln-pdc-adc-dpc-hcs}.
 It is observed that for both the quantum correlation measures, the local depolarizing channel is maximally destructive. The local amplitude damping channel is much less destructive in both cases. 
 The local phase damping channel has, however a richer behavior. It is minimally destructive to logarithmic negativity, while surprisingly being almost maximally destructive to quantum discord.
 \section{GHZ state under decoherence} 
 \label{ghz-decoh}
Up to now, we have studied decoherence effects on the  \(H_{C_{N}}^{m}\) state for different models of decoherence. In this section, we study effects of these noise models 
on the well-known macroscopic state, the GHZ state.  The GHZ state, consisting of \(k\) parties in the  microscopic part and \(N\) in the macroscopic one, is given by 
 \label{eq:ghz-state}
 \begin{align}
  \arrowvert \mbox{GHZ}\rangle_{N+k}&=\frac{1}{\sqrt{2}}(\arrowvert 0^{\otimes k}\rangle_{\mu}\otimes\arrowvert 0^{\otimes N}\rangle_{M}\nonumber\\
  &+\arrowvert 1^{\otimes k}\rangle_{\mu}\otimes\arrowvert 1^{\otimes N}\rangle_{M}).
 \end{align}
Just like the \(H_{C_{N}}^{m}\) state, the \(\arrowvert\mbox{GHZ}\rangle_{N+k}\) state also possesses one ebit of entanglement which is the same as its quantum discord, in the microscopic to
macroscopic bipartition. In this section, we will study the trends of quantum correlations of the   \(\arrowvert \mbox{GHZ}\rangle_{N+k}\)
state, against the three local noise models that we had
considered for the \(H_{C_{N}}^{m}\) state,  and compare with those of the   \(H_{C_{N}}^{m}\) state.

\subsection{GHZ state under local phase damping channel}
 Let us begin by considering  the effect of local phase damping channels on the  GHZ state. 
After each qubit of the GHZ state is sent through a  phase damping  channel, the resulting state  can be written as
\begin{align}
 \rho_{\mbox{GHZ}_{N+k}}^{lpdc}&=\frac{1}{2}\Big(P_{0}^{\mu \otimes k}\otimes P_{0\ldots0}+P_{1}^{\mu\otimes k}\otimes P_{1\ldots 1}\nonumber \\
 &+(1-2\beta)^{N+k}(P_{0,1}^{\mu\otimes k}\otimes P_{0\ldots 0,1\ldots1}+h.c.)\Big),
\end{align}
where   \(\beta=\frac{p}{2}\), 
\(P_{0}^{\mu}=\arrowvert 0\rangle\langle 0\arrowvert\),  
\(P_{1}^{\mu}=\arrowvert 1\rangle\langle 1\arrowvert\),
\(P_{0,1}^{\mu}=\gamma\arrowvert 0\rangle\langle 1\arrowvert\),
\(P_{0\ldots 0}=\arrowvert 0\ldots 0\rangle\langle 0\ldots 0\arrowvert\),  
\(P_{1\ldots 1}=\arrowvert 1\ldots 1\rangle\langle 1\ldots 1\arrowvert\),  
\(P_{0\ldots 0,1\ldots 1}=\arrowvert 0\ldots 0\rangle\langle 1\ldots 1\arrowvert\). 
Here, $\gamma=1-p$. After performing  the partial transpose on the state  \(\rho_{\mbox{GHZ}_{N+k}}^{lpdc}\) with respect to micro : macro bipartition, the  matrix \(B_{\mbox{GHZ}_{N+k}}^{lpdc}\), whose eigenvalues 
contribute to the entanglement of the noisy GHZ\(_{N+k}\) state in the micro : macro bipartition, is found to be of the form
\[B_{\mbox{GHZ}_{N+k}}^{lpdc}= \left( \begin{array}{cc}
0 &  \frac{(1-2\beta)^{N+k}}{2} \\
\frac{(1-2\beta)^{N+k}}{2}&0 \end{array} \right),\]
and hence the entanglement is given by
\begin{equation}
\label{eq:ln-ghz-pd}
 E_{\mbox{GHZ}_{N+k}}^{lpdc}=\log_{2}[(1-p)^{N+k}+1].
\end{equation}
It is clear from  Eq. (\ref{eq:ln-ghz-pd}) that the entanglement of the  state depends on the number of the particles in the macroscopic part. This is in sharp contrast to the situation 
in the case of the \(H_{C_{N}}^{m}\) state,  where quantum correlations do not depend on the number of particles in the macroscopic part of the state. Note that the exponential of the 
logarithmic negativity decreases as  \((1-p)^{k+m}\) for the \(H_{C_{N}}^{m}\) state (Eq. (\ref{eq:lpdcnm})),
 while as \((1-p)^{N+k}\) for the GHZ  state  (Eq. (\ref{eq:ln-ghz-pd})). 
Since \(m<N\), the entanglement of the \(H_{C_{N}}^{m}\) state under a noisy environment, as modeled by local dephasing channels, is more than that of the \(\mbox{GHZ}_{N+k}\) state, for any \(N\), \(m\), and \(k\). 
The entanglements match for $m=N$. The \(H_{C_{N}}^{m}\) and  \(\mbox{GHZ}_{N+k}\) states are equal up to local unitary transformations for $m=N$. But this does not imply equality of the entanglements for all channels,
as the local unitary transformations and the local decohering channels may 
not commute. Note also that it is certainly possible to consider the case when $1\leq m\ll N$, and in such cases, $(1-p)^{k+m}$ and $(1-p)^{k+N}$ are very different.

\subsection{GHZ state under local amplitude damping channel}
\begin{figure}
 \includegraphics[angle=270,width=4.2cm,totalheight=3.7cm]{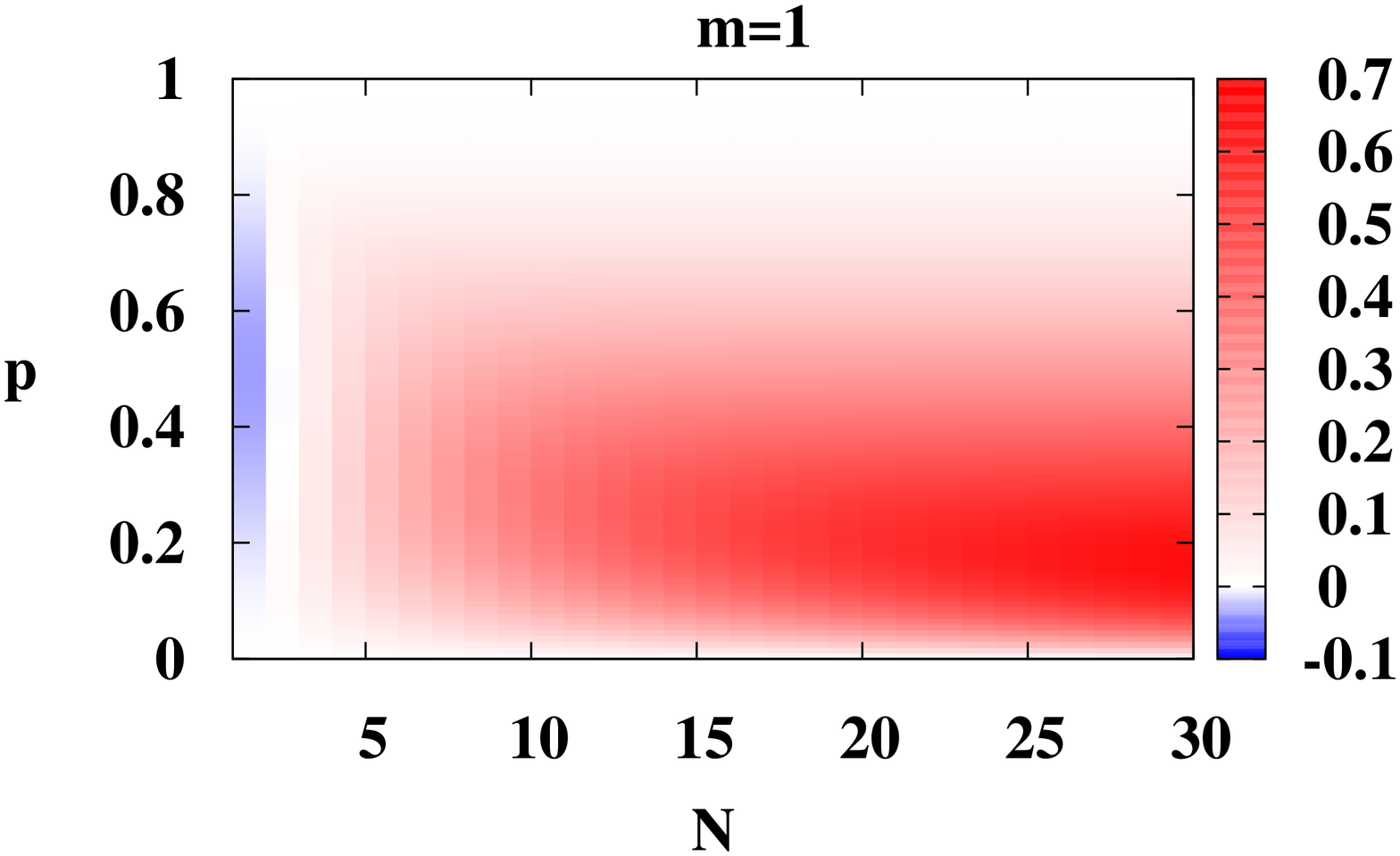}
 \includegraphics[angle=270,width=4.2cm,totalheight=3.7cm]{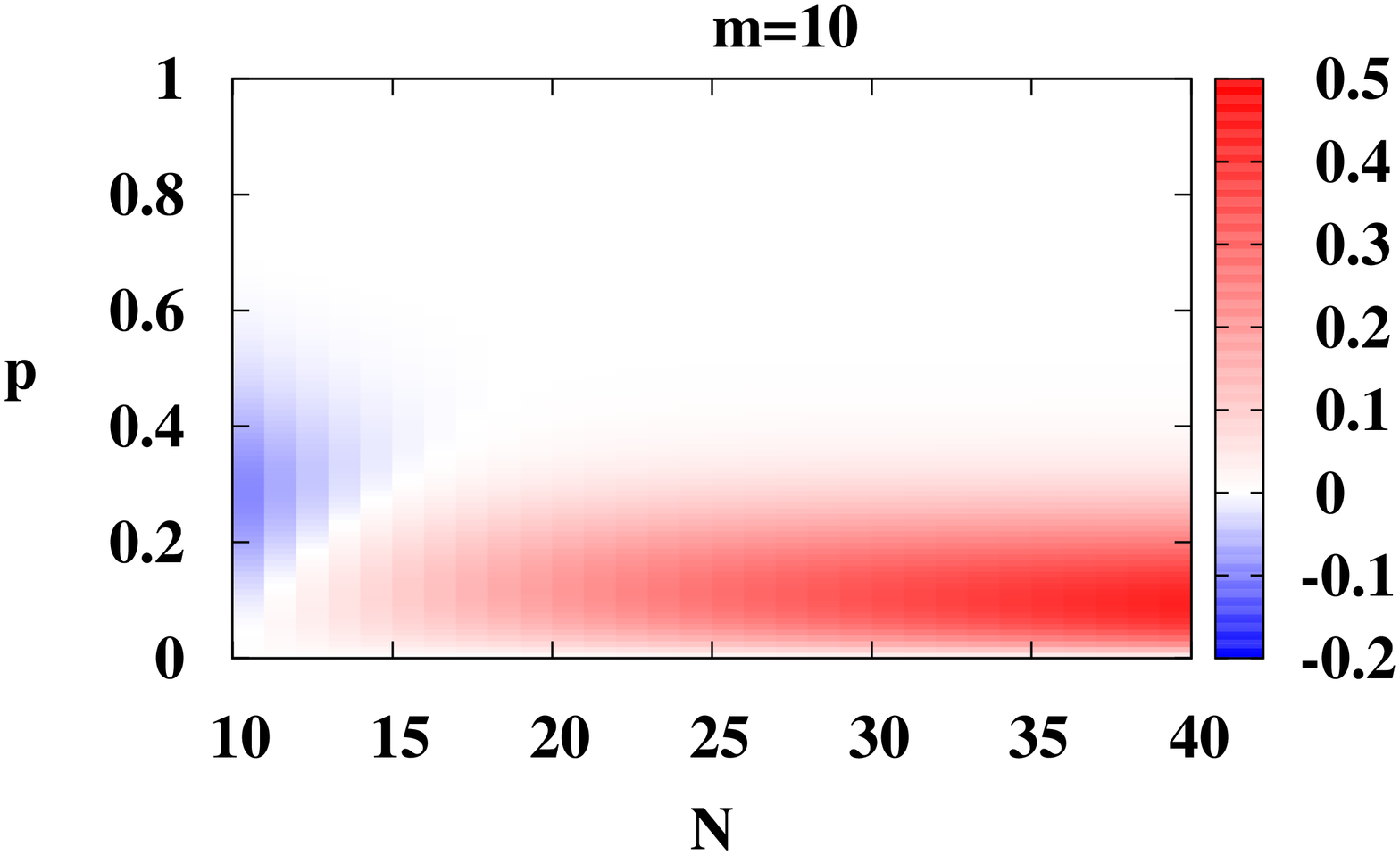}\\
 \includegraphics[angle=270,width=4.2cm,totalheight=3.7cm]{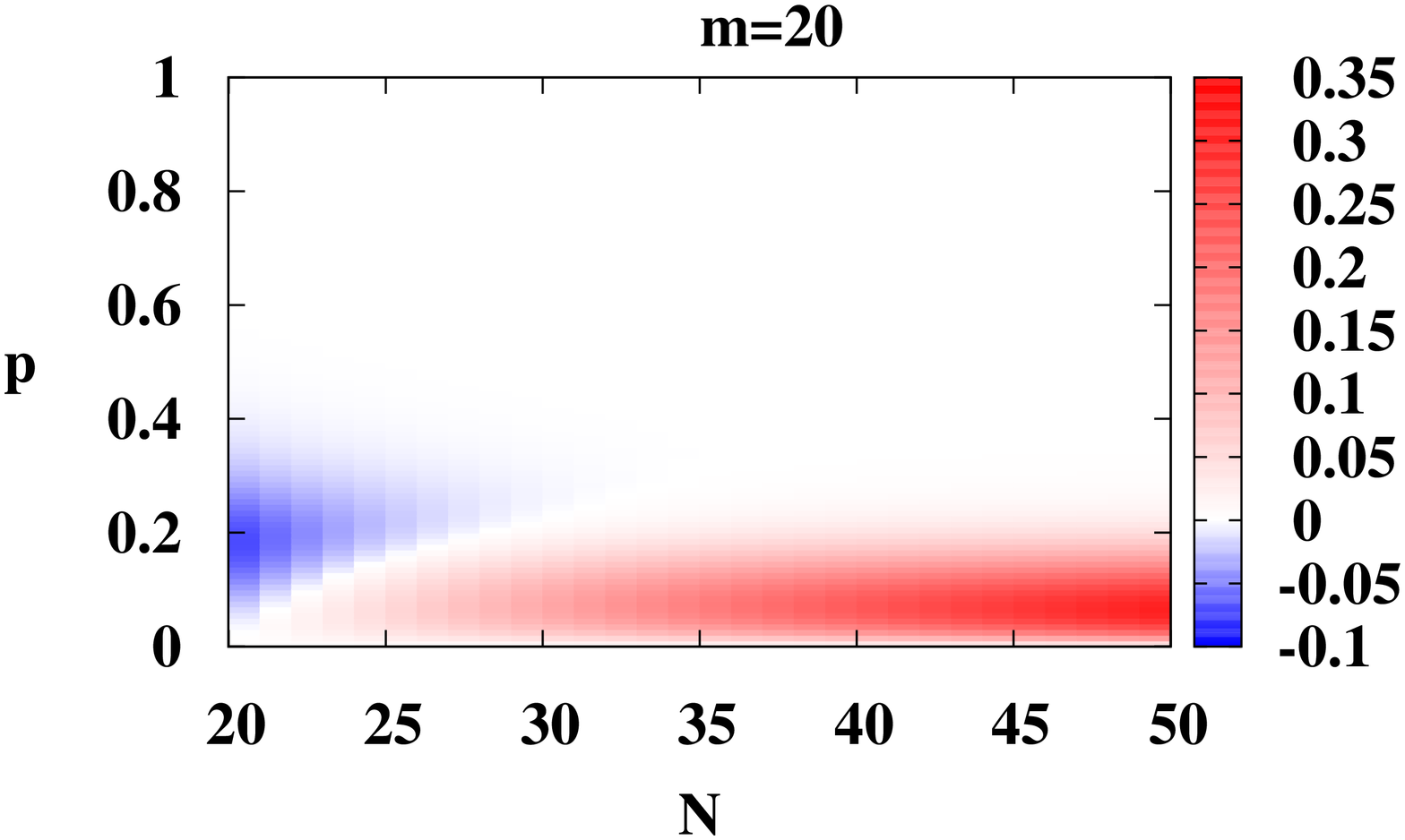}
 \includegraphics[angle=270,width=4.2cm,totalheight=3.7cm]{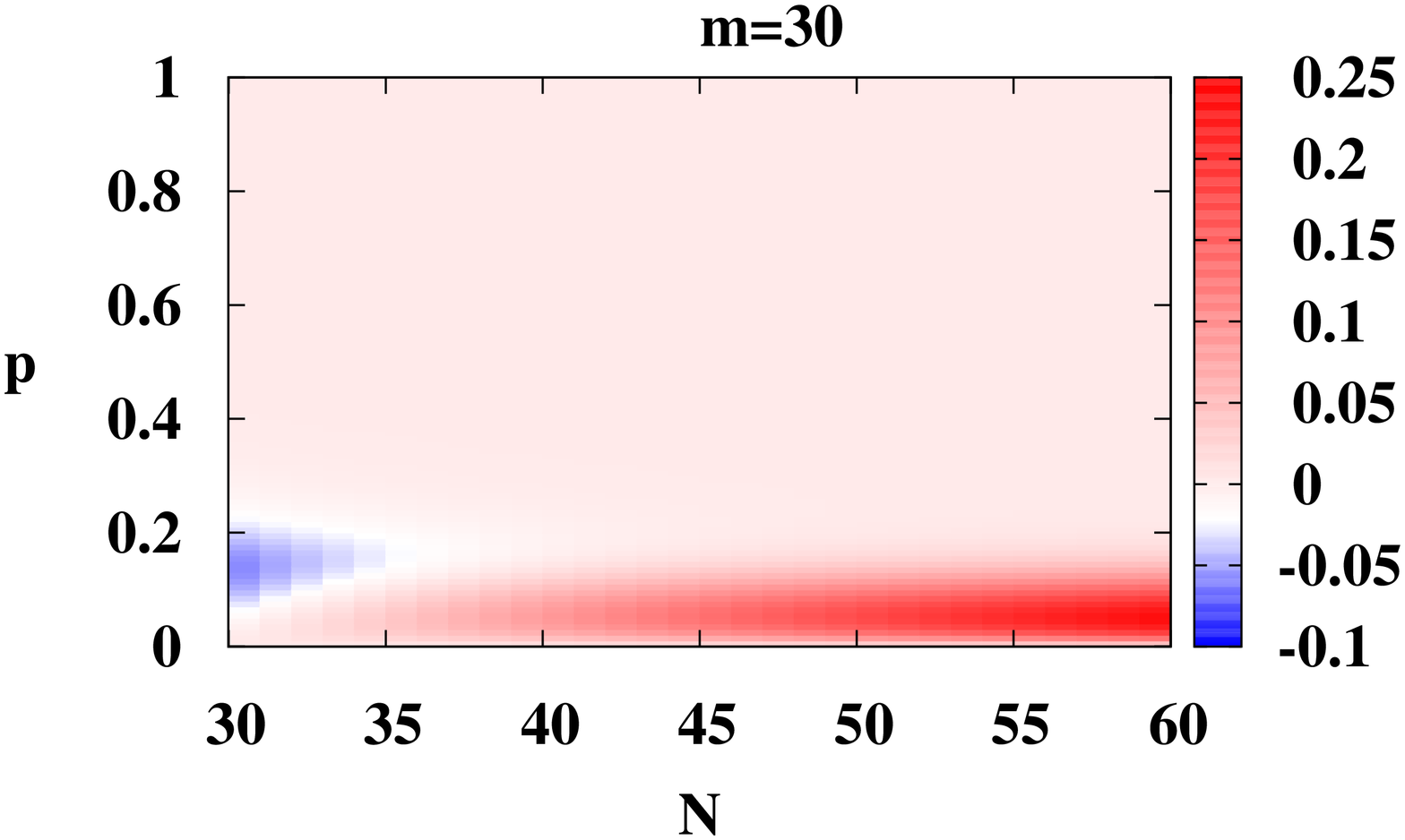}
 \caption{(Color) Comparison of the entanglements of the local amplitude damped 
 \(H_{C_{N}}^{m}\) and $\mbox{GHZ}_{N+k}$ states. We plot here the projections of the difference between the logarithmic 
 negativities (logarithmic negativity of the noisy \(H_{C_{N}}^{m}\) state minus that of the noisy $\mbox{GHZ}_{N+k}$ state) of the local amplitude damped \(H_{C_{N}}^{m}\) and $\mbox{GHZ}_{N+k}$ states for  $k=1$ and $m=1,10,20,30$,  against the amplitude damping parameter, $p$, and the number of
 qubits, \(N\), in the macroscopic part. The logarithmic negativities are measures in ebits, while \(N\) is measured in qubits. \(p\) 
 is a dimensionless parameter.}
 \label{fig:ln-adc-ghz-hcnm1}
 \end{figure}

\begin{figure}
 \includegraphics[angle=270,width=4.2cm,totalheight=3.7cm]{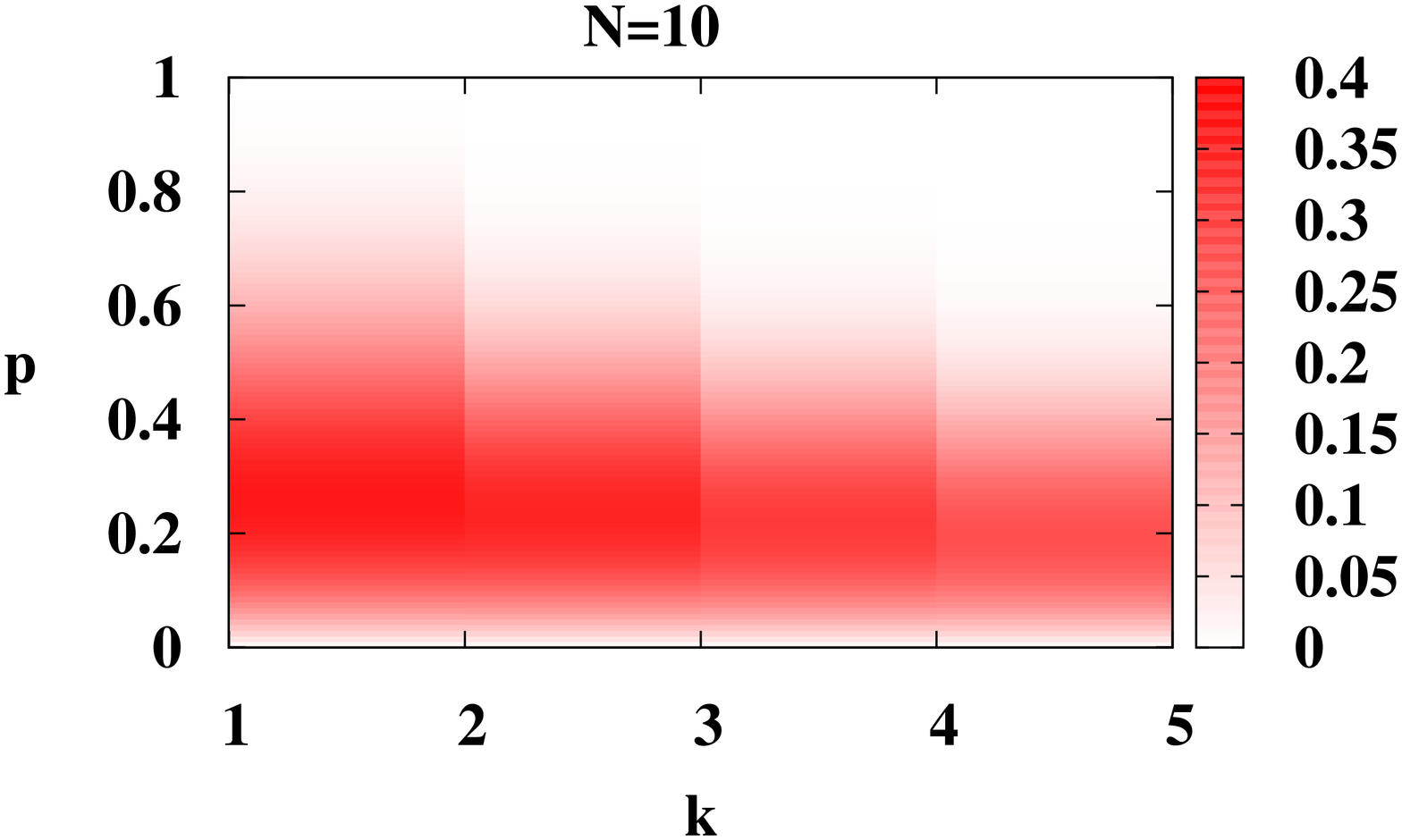}
  \includegraphics[angle=270,width=4.2cm,totalheight=3.7cm]{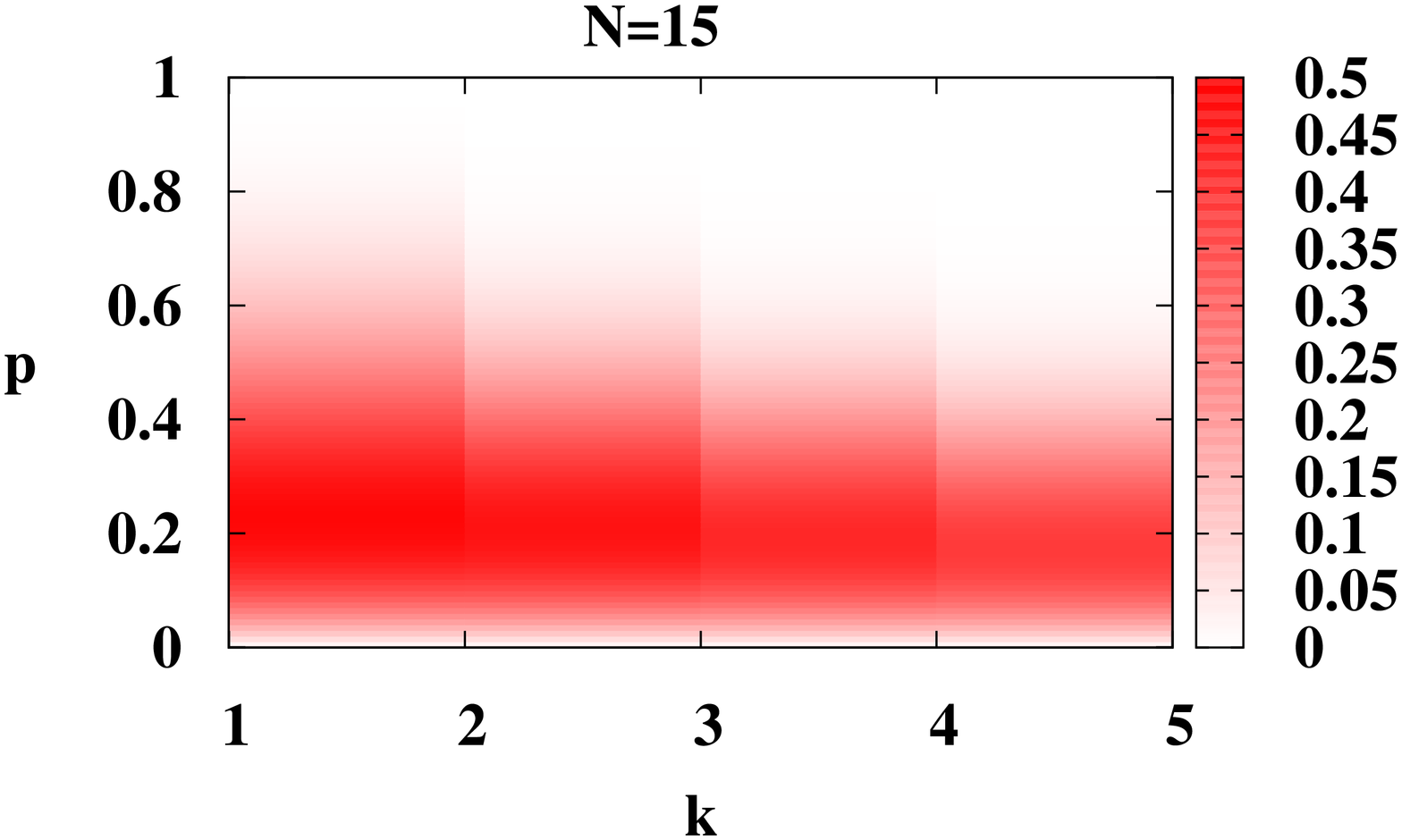}\\
  \includegraphics[angle=270,width=4.2cm,totalheight=3.7cm]{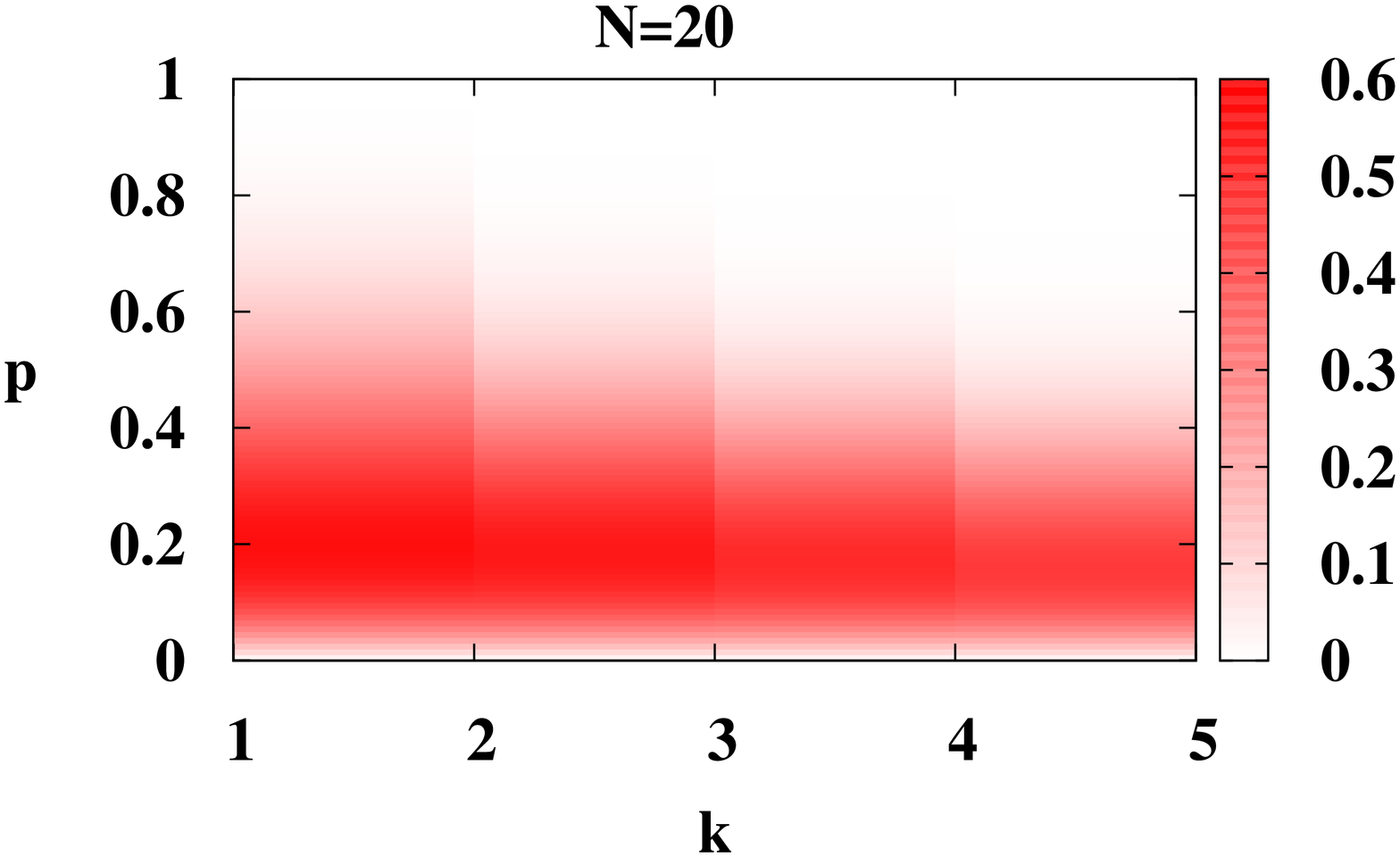}
  \includegraphics[angle=270,width=4.2cm,totalheight=3.7cm]{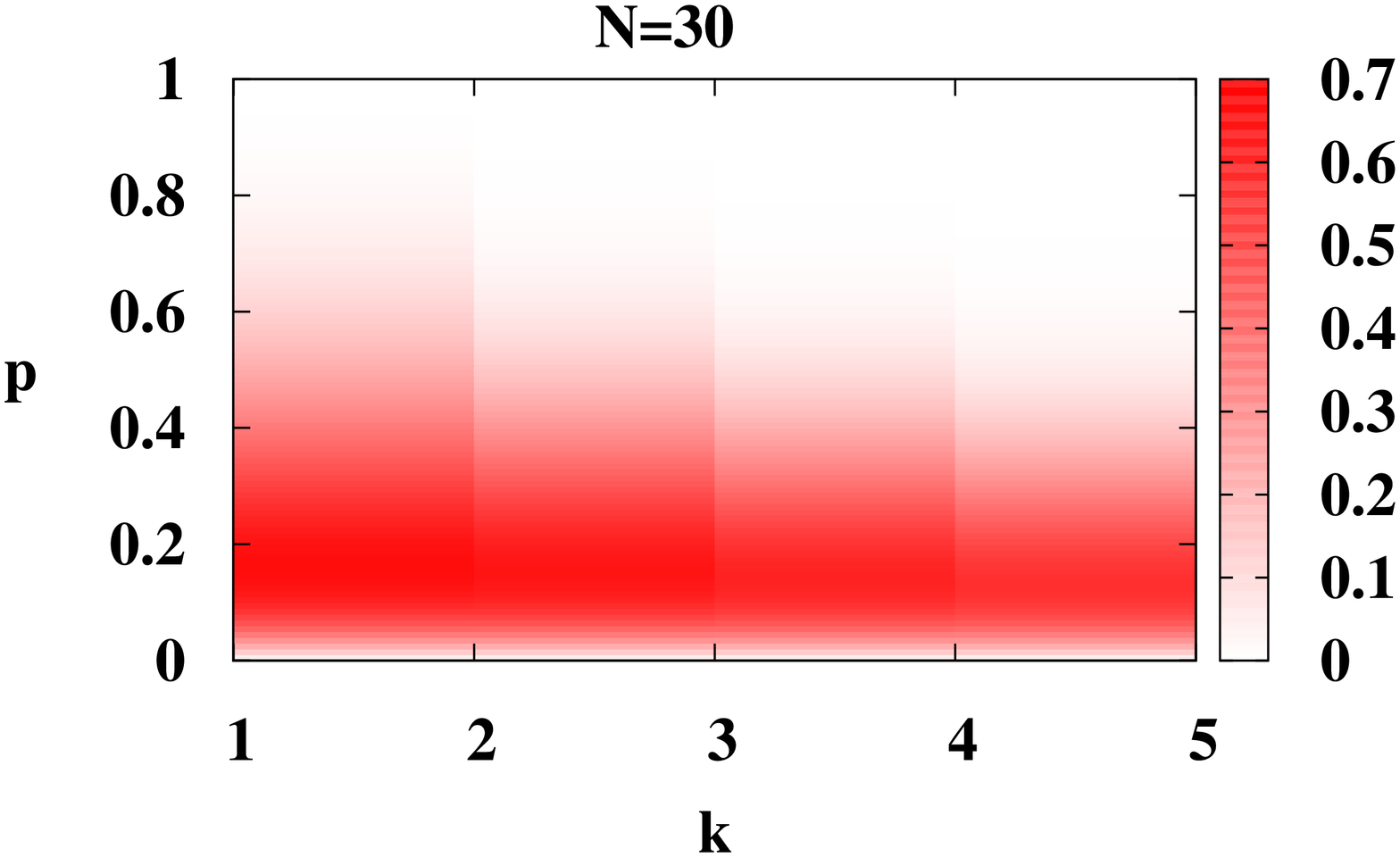}
\caption{(Color) 
A further comparison of the entanglements of the local amplitude damped 
 \(H_{C_{N}}^{m}\) and $\mbox{GHZ}_{N+k}$ states. We plot here the projections of the difference between the logarithmic 
 negativities (logarithmic negativity of the noisy \(H_{C_{N}}^{m}\) state minus that of the noisy $\mbox{GHZ}_{N+k}$ state) of the local amplitude damped \(H_{C_{N}}^{m}\) and $\mbox{GHZ}_{N+k}$ states for  $m=1$ and $N=10,15,20,30$,  against the amplitude damping parameter, $p$, and the number of
 qubits, \(k\), in the microscopic part. The logarithmic negativities are measures in ebits, while \(k\) is measured in qubits. \(p\) 
 is a dimensionless parameter.
%
}
 \label{fig:ln-adc-ghz-hcnm1-k}
 \end{figure}

In this case, all the qubits of the \(\mbox{GHZ}_{N+k}\) state are sent through  amplitude damping channels. 
%
The block of the total matrix, after partial transposition, which is responsible for non-zero logarithmic negativity is
\[B_{\mbox{GHZ}_{N+k}}^{ladc}= \frac{1}{2}\left( \begin{array}{cc}
p^{k}\gamma^{N} &  \gamma^{\frac{N+k}{2}} \\
\gamma^{\frac{N+k}{2}}&p^{N}\gamma^{k}  \end{array} \right).\]
Correspondingly, the eigenvalue which can be negative for some values of $p$, is given by
\begin{align}
 \lambda_{\mbox{GHZ}_{N+k}}^{ladc}&=\frac{1}{4}\Big(p^{k}\gamma^{N}+p^{N}\gamma^{k} \nonumber \\
 &-\sqrt{(p^{k}\gamma^{N}+p^{N}\gamma^{k})^{2}+4\gamma^{N+k}(1-p^{N+k})}\Big).
 \label{eq:neg-ghz-ampdc}
\end{align}
For fixed \(N\), comparing Eqs. (\ref{eq:hcs-ampdc}) and (\ref{eq:neg-ghz-ampdc}), we find that 
under the local amplitude damping channel,
the \(H_{C_{N}}^{1}\) state has higher entanglement  than the \(\mbox{GHZ}_{N+k}\)
state for all $N >2$ and for  \(k\ll N\).  
See Figs. \ref{fig:ln-adc-ghz-hcnm1} (top left panel) and \ref{fig:ln-adc-ghz-hcnm1-k}. 
Moreover, the ``critical value'' of the decohering parameter, \(p\), at which the decohered state becomes 
separable, is always greater for the \(H_{C_{N}}^{1}\) state than that for the \(\mbox{GHZ}_{N+k}\) state 
for all \(N>2\) and for $k \ll N$ (see also Table I).
For higher values of \(m\), i.e. for \(m>1\), the \(H_{C_{N}}^{m}\) state has higher entanglement than the GHZ state, provided we 
choose sufficiently high \(N\). See Figs. \ref{fig:ln-adc-ghz-hcnm1}.
Note that unlike  the case of the local phase damping channel, the entanglements of the local amplitude damped \(H_{C_{N}}^{N}\) and  \(\mbox{GHZ}_{N+k}\) states do not match. However, 
the local amplitude damped \(H_{C_{N}}^{N}\) can have only a lower entanglement than 
the local amplitude damped \(\mbox{GHZ}_{N+k}\). This is clearly seen from Eqs. (\ref{eq:hcs-ampdc}) (for $m=N$) and (\ref{eq:neg-ghz-ampdc}) (see also Fig. \ref{fig:ln-adc-ghz-hcnm1} for $m=1,10,20,30$). 
%

\subsection{GHZ state under local depolarizing channel}
\begin{figure}
 \includegraphics[angle=270,width=4.2cm,totalheight=3.7cm]{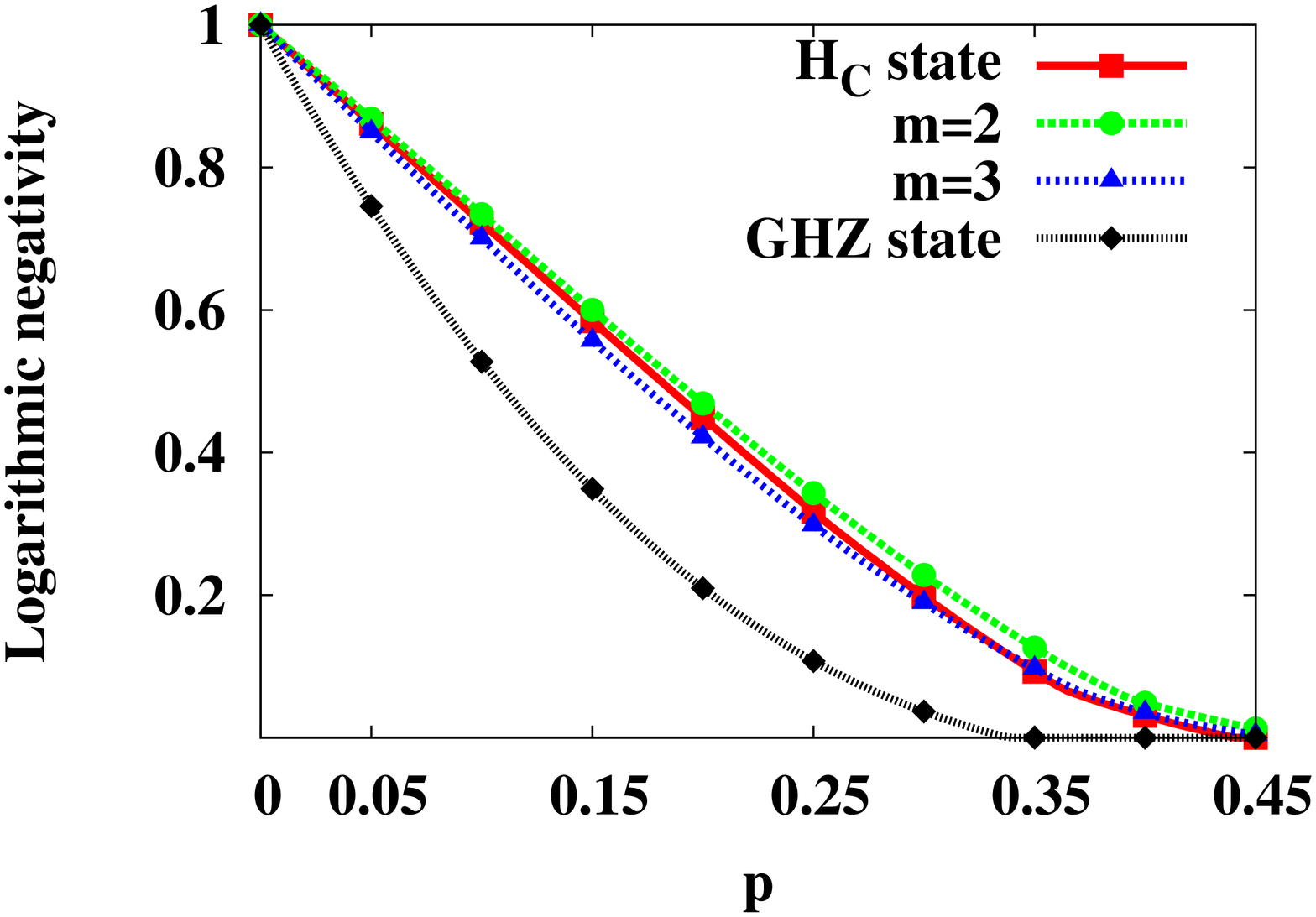}
 \includegraphics[angle=270,width=4.2cm,totalheight=3.7cm]{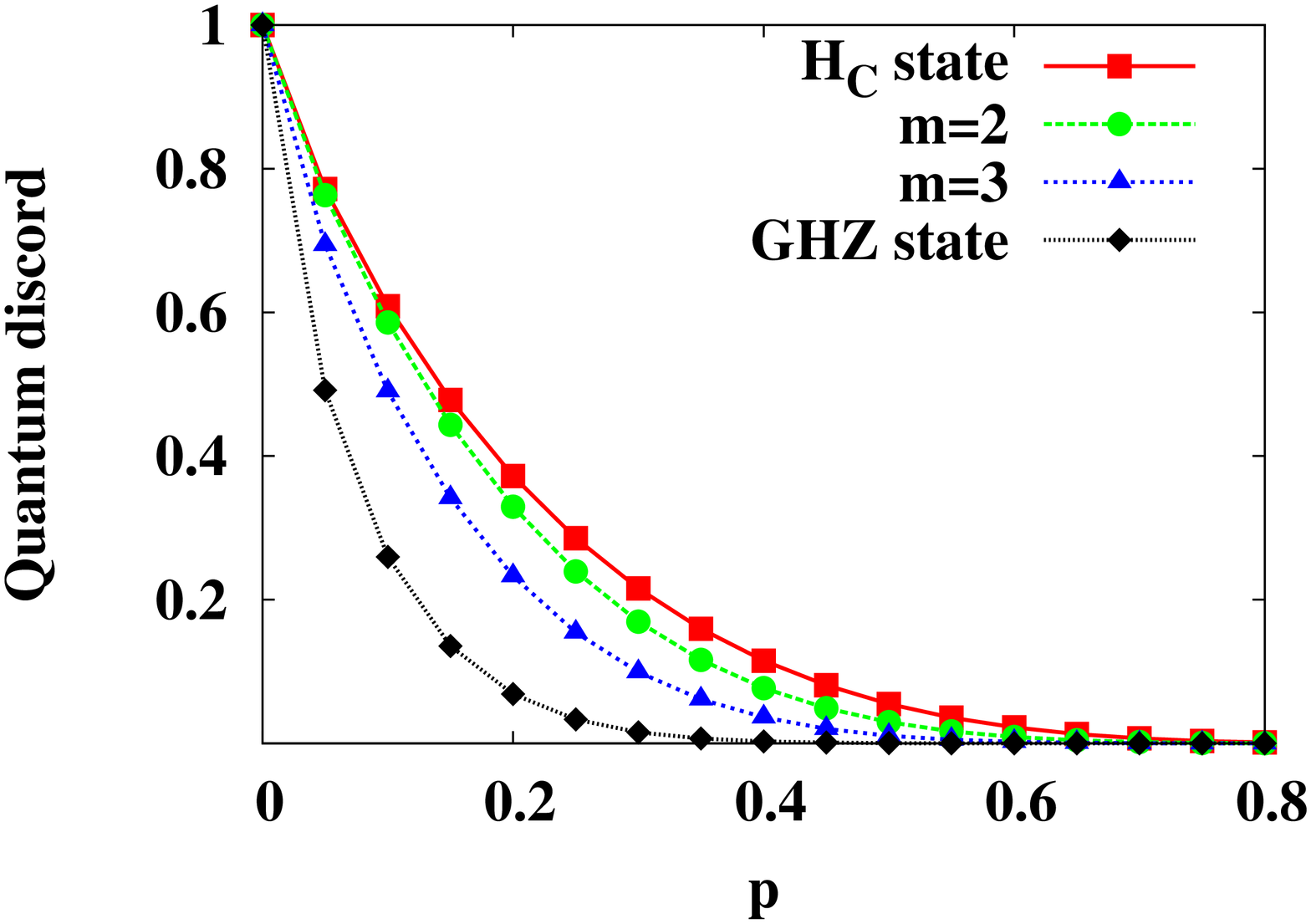}
\caption{(Color online) The panels are the same as in Fig. \ref{fig:ln-dpc-hcs}, except that there is
an additional curve in each panel corresponding to the GHZ state for \(k=1\), \(N=6\).
See Table I for numerical values.}
 \label{fig:ln-dpc-ghz-HCS}
 \end{figure}

\begin{figure}
 \includegraphics[angle=270,width=6.2cm,totalheight=3.7cm]{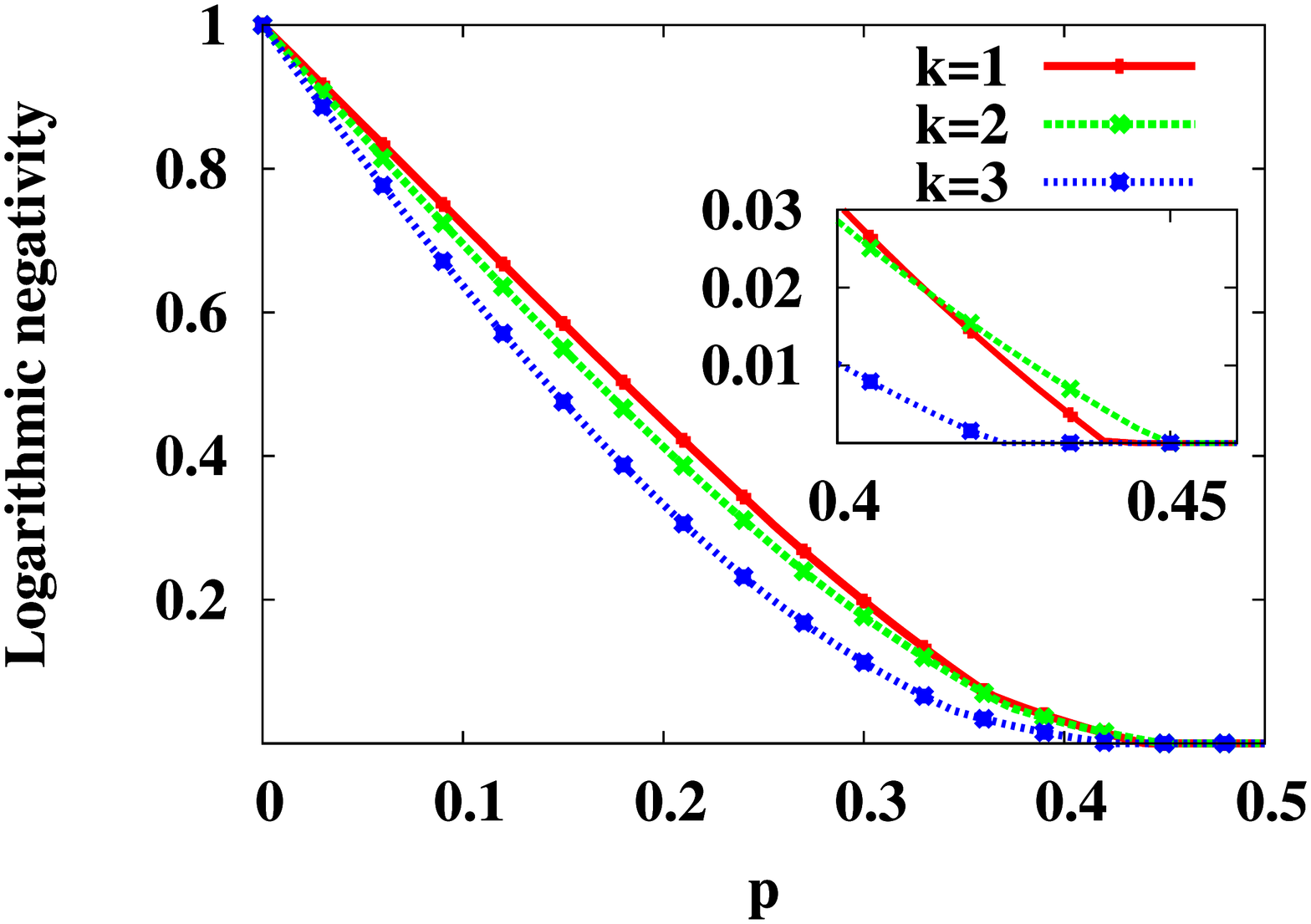}
  \includegraphics[angle=270,width=6.2cm,totalheight=3.7cm]{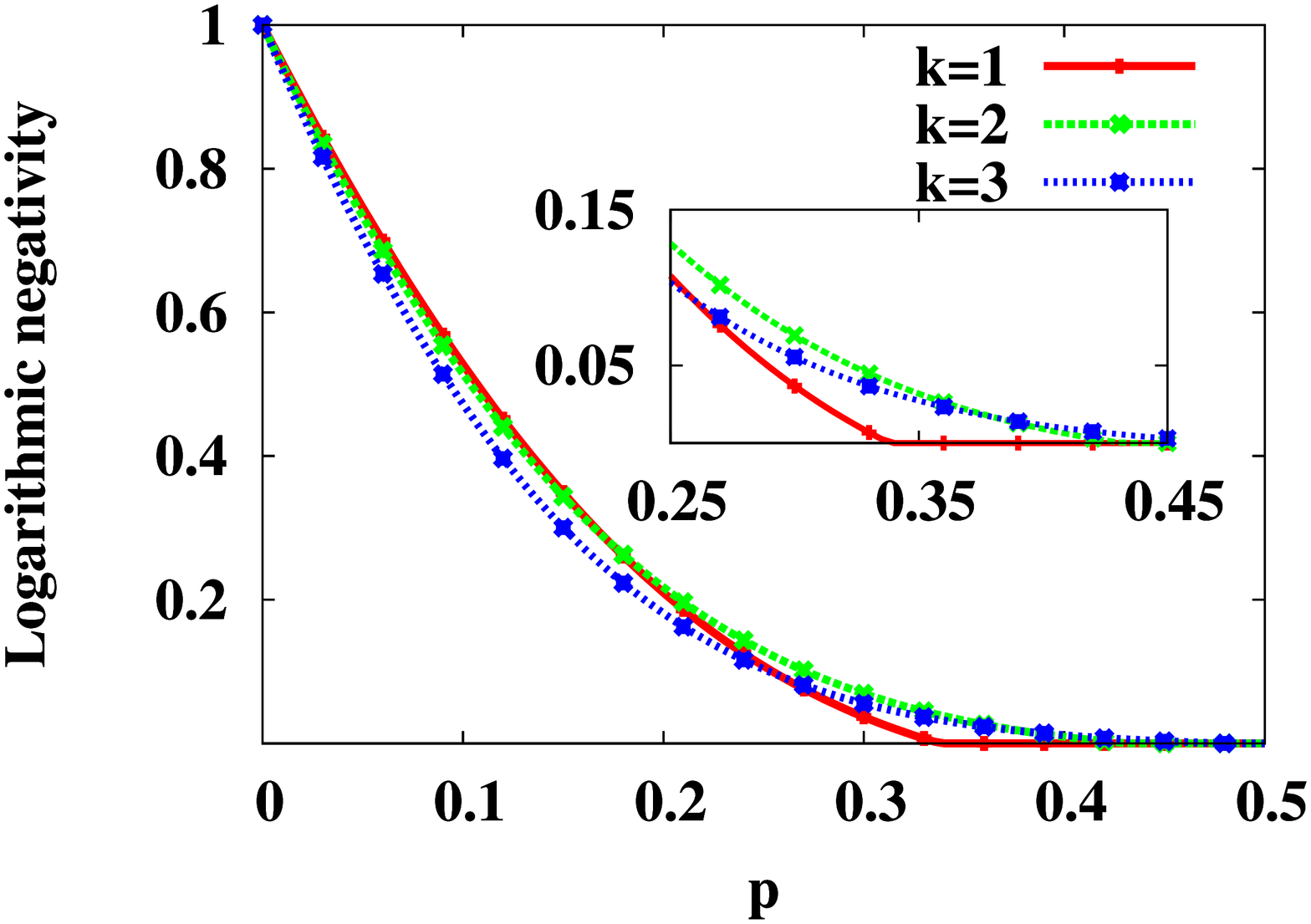}
 \caption{(Color online) \(H_{C_{N}}^{1}\) (top panel) versus and GHZ (bottom panel) after being affected by local depolarizing channels for different number of particles in the macroscopic sectors. The vertical axes represent logarithmic negativity (in ebits) in the 
 micro : macro bipartition, the horizontal axes correspond to the depolarizing parameter, \(p\) (dimensionless). We choose \(N=6\). 
 As is clear from the bottom panel, increase of \(k\) for a fixed \(N\) leads to increase in logarithmic negativity for the GHZ state. The 
 insets reveal the situations where the entanglements vanish. The axes of the insets represent the same quantities as of the 
 corresponding parent figures.
}
 \label{fig:ln-dpc-hcs-diff-k}
 \end{figure}

\begin{figure}
 \includegraphics[angle=270,width=4.2cm,totalheight=3.7cm]{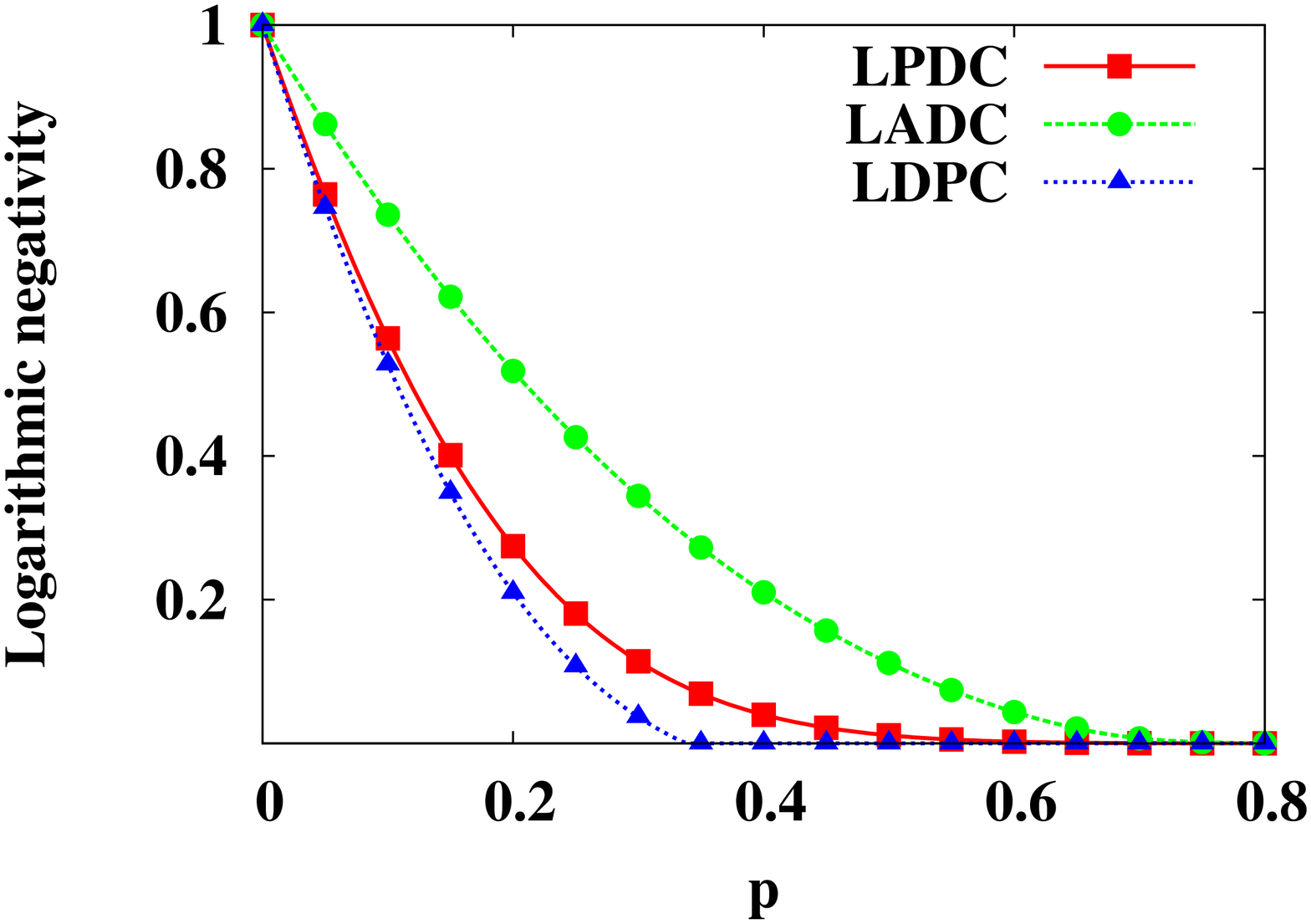}
  \includegraphics[angle=270,width=4.2cm,totalheight=3.7cm]{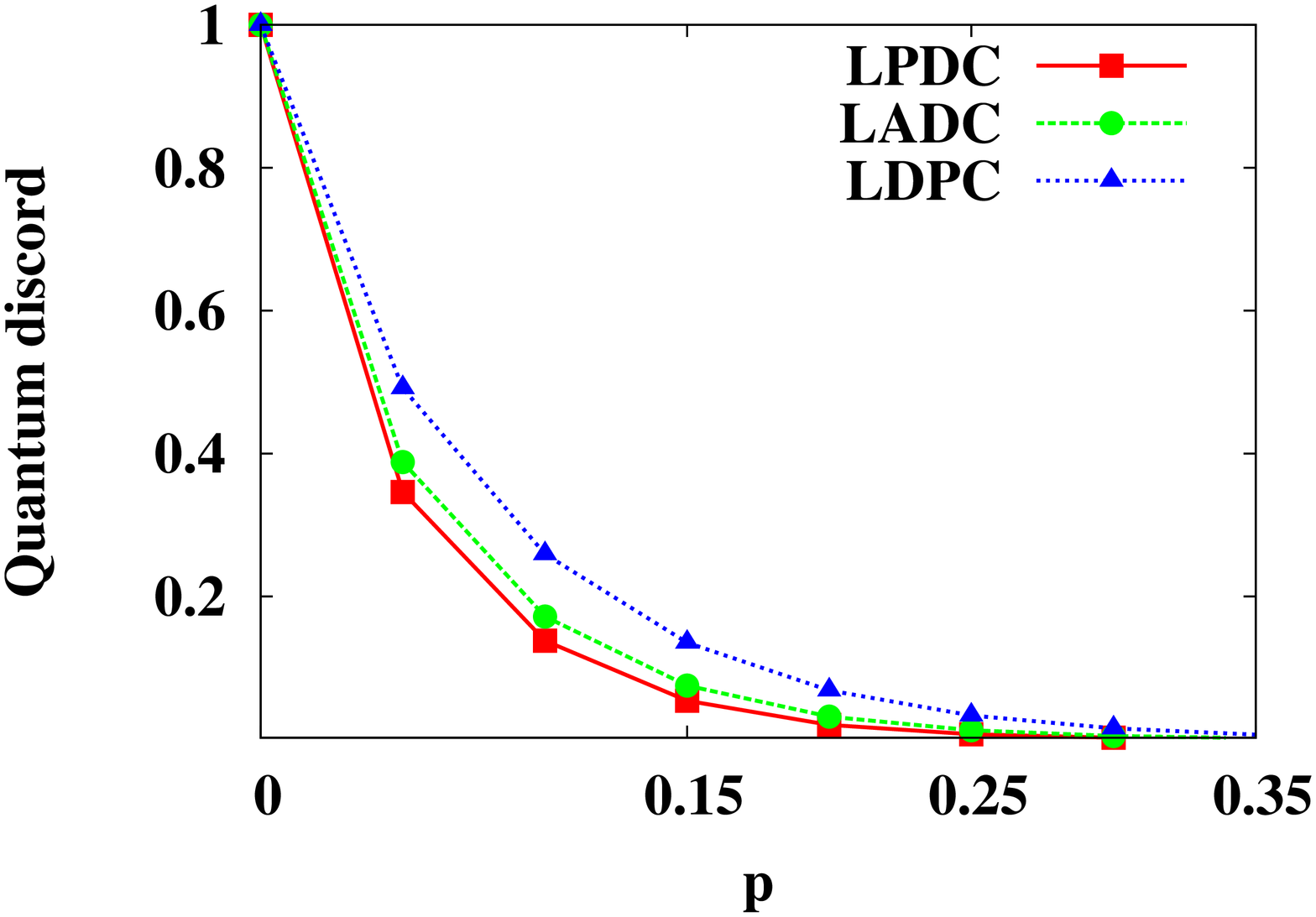}
\caption{(Color online) The panels are the same as in Fig. \ref{fig:ln-pdc-adc-dpc-hcs}, except that all the curves correspond now to the 
noise-affected GHZ states instead of the \(H_{C_{N}}^{1}\) states. Comparing the panels here with those 
of Fig. \ref{fig:ln-pdc-adc-dpc-hcs}, we see that the noise-affected GHZ states behave very differently from the corresponding 
noise-affected \(H_{C_{N}}^{1}\) states. This is especially true for the entanglements of the 
local phase damped states and for the quantum discords of the states after being affected by any type of noise. 
}
 \label{fig:ln-adc-pdc-dpc-ghz}
 \end{figure}

Let us  now study the effect of local depolarizing channels on the  \(\mbox{GHZ}_{N+k}\) state. To calculate the entanglement, in the micro : macro bipartition, we have to find the  
negative eigenvalue of the  matrix given by
\[B_{\mbox{GHZ}_{N+k}}^{ldpc}= \frac{1}{2}\left( \begin{array}{cc}
a &  b \\
b&a \end{array} \right),\]
where 
\(a=\alpha^{k} \beta^{N}+\beta^{k} \alpha^{N}\), \(b=\gamma^{N+k}\), \(\gamma=(1-p)\) and  \(\alpha=(1-\frac{p}{2})\) .
The eigenvalue of \(B_{\mbox{GHZ}_{N+k}}^{ldpc}\), which is negative for some values of $p$, is given by
\begin{equation}
 \lambda_{\mbox{GHZ}_{N+k}}^{ldpc}=\frac{1}{2}[\alpha^{k}\beta^{N}+\alpha^{N}\beta^{k}-\gamma^{N+k}].
\end{equation}
The logarithmic negativity, then, is given by
\begin{equation}
 E_{\mbox{GHZ}_{N+k}}^{ldpc}=\mbox{log}_2(2|\mbox{min}(0,\lambda_{\mbox{GHZ}_{N+k}}^{ldpc})|+1).
\end{equation}
For six particles in the macroscopic sector and a single particle in the microscopic one, a comparison of entanglement and discord between  the \(\arrowvert H_{C_{N}}^{1}\rangle\), 
\(\arrowvert H_{C_{N}}^{2}\rangle\), \(\arrowvert H_{C_{N}}^{3}\rangle\), and the GHZ states, after they are affected by local depolarizing channels, is presented in Fig. \ref{fig:ln-dpc-ghz-HCS}. 
We find that the  \(\arrowvert H_{C_{N}}^{1}\rangle\), \(\arrowvert H_{C_{N}}^{2}\rangle\) and \(\arrowvert H_{C_{N}}^{3}\rangle\) can sustain about \(43\%\), \(46\%\), and \(45\%\) noises respectively,
while the GHZ remains robust against up to \(34\%\) noise, for \(k=1\) and \(N=6\). When \(k=2\), the value of \(p\) at which logarithmic negativity vanishes is \(0.445\) for the \(\arrowvert H_{C_{N}}^{1}\rangle\)
state while it is \(0.43\) for the GHZ state. When \(k=3\) these values are \(0.42\) and \(0.48\) for the \(\arrowvert H_{C_{N}}^{1}\rangle\) and GHZ states, respectively.
A comparison between the \(\arrowvert H_{C_{N}}^{1}\rangle\) and GHZ states, after they are affected by local depolarizing channels, is presented in Fig. \ref{fig:ln-dpc-hcs-diff-k} for different values of $k$. A comparison with respect to robustness of
entanglement and quantum discord, between the different noise models, of the GHZ state is presented in Fig. \ref{fig:ln-adc-pdc-dpc-ghz}. The latter comparison reveals a picture that is quite different
from that obtained in a similar comparison in Fig. \ref{fig:ln-pdc-adc-dpc-hcs} for the $H_{C_{N}}^{m}$ state. See Figs. \ref{fig:ln-pdc-adc-dpc-hcs} and Fig. \ref{fig:ln-adc-pdc-dpc-ghz} for more details.
\begin{table}[ht]
Table for critical values of decoherence parameter for different states under different noisy channels.
\centering
\begin{tabular}{c c c c c c c c}
\hline\hline
Channel & \(H_{C_{N}}^{1}\) & \(H_{C_{N}}^{2}\) & \(H_{C_{N}}^{3}\) & GHZ & \(G\) &\(H_{C_{N}}^{N-1}\) &\(H_{C_{N}}^{N}\) \\ [0.5ex]
\hline
LPDC & 97.5 & 92 & 86 & 67.5 & 81 & 73 & 67.5 \\
LADC & 97 & 91 & 84 & 78 & 87 & 73.5 & 75\\
LDPC & 43 & 46 & 45 & 33 & 43 & 39 & 33\\
\hline
\end{tabular}
\label{table:nonlin}
\caption{Comparison of ``critical values''
 of decoherence at which the entanglement vanishes for different states under different noisy channels. The critical point has been taken at the value of decoherence parameter where 
logarithmic negativity becomes less or equal to \(10^{-4}\).
The values exhibited as percentages of the noise level that the corresponding state can sustain before becoming separable.
 The states considered to construct this Table consists of one qubit in its microscopic sector and six qubits in its macroscopic 
sector. The first column shows the type of local noise acting on the state.
 The second, third, fourth, fifth, sixth, seventh, and eighth 
columns exhibit the critical values for the 
 \(H_{C_{N}}^{1}\), \(H_{C_{N}}^{2}\) , \(H_{C_{N}}^{3}\), GHZ, \(G\), \(H_{C_{N}}^{N-1}\), 
and \(H_{C_{N}}^{N}\) states respectively. 
 The \(G\) state appears in Eq. (\ref{eq:state-1-pd}). Note that the critical values for the GHZ and \(H_{C_{N}}^{N}\) states
 are the same for the  LPDC and LDPC, while they differ for the LADC.}
\end{table}
 \section{Other macroscopic states}
 \label{other-macro-states}
 In this section, we will discuss the effects of noise on quantum correlations of some further  macroscopic states. We then compare their results with those of the  \(H_{C_{N}}^{m}\) and \(\mbox{GHZ}_{N+k}\) states.
 \subsection{\(G\) state}
 Consider the  state
 \begin{equation}
 \label{eq:state-1-pd}
 \arrowvert G \rangle _{\mu A_{1}...A_{N}} = \frac{1}{\sqrt{2}}\left(\arrowvert 0 \rangle _{\mu} \arrowvert W_{N} \rangle _{A_{1}...A_{N}}
+
\arrowvert 1 \rangle _{\mu} \arrowvert \widetilde {W} \rangle_{A_{1}...A_{N}}\right),
\end{equation}
introduced in \cite{Gdansk}, where 
\begin{equation}
 \arrowvert \widetilde W_{N} \rangle _{A_{1}...A_{N}} = \sigma_{x}^{\otimes N} \arrowvert W_{N} \rangle _{A_{1}...A_{N}}, 
\end{equation}
with \(\sigma_x = |0\rangle \langle 1| + |1\rangle \langle 0|\).
This state is a cat-like state in the sense, that the states \(|W\rangle\) and \(|\widetilde{W}\rangle\) are macroscopically distinct in terms of their 
\(\sigma_z\)-magnetizations, similar to the case of the GHZ state.
In the absence of noise, this state also possesses maximum entanglement in the micro : macro bipartition.
 
\begin{figure}
\includegraphics[angle=270,width=4.2cm,totalheight=3.7cm]{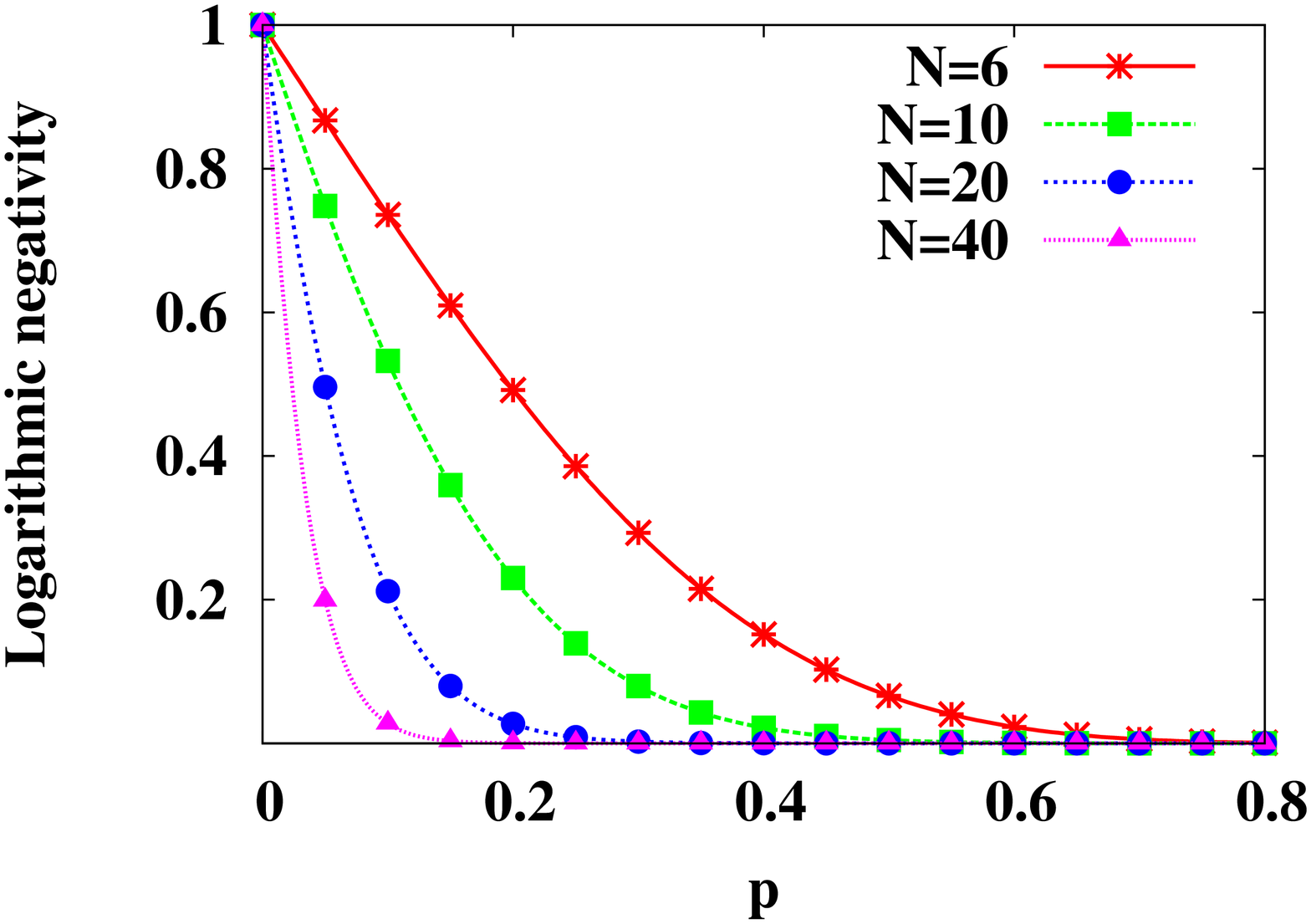}
\includegraphics[angle=270,width=4.2cm,totalheight=3.7cm]{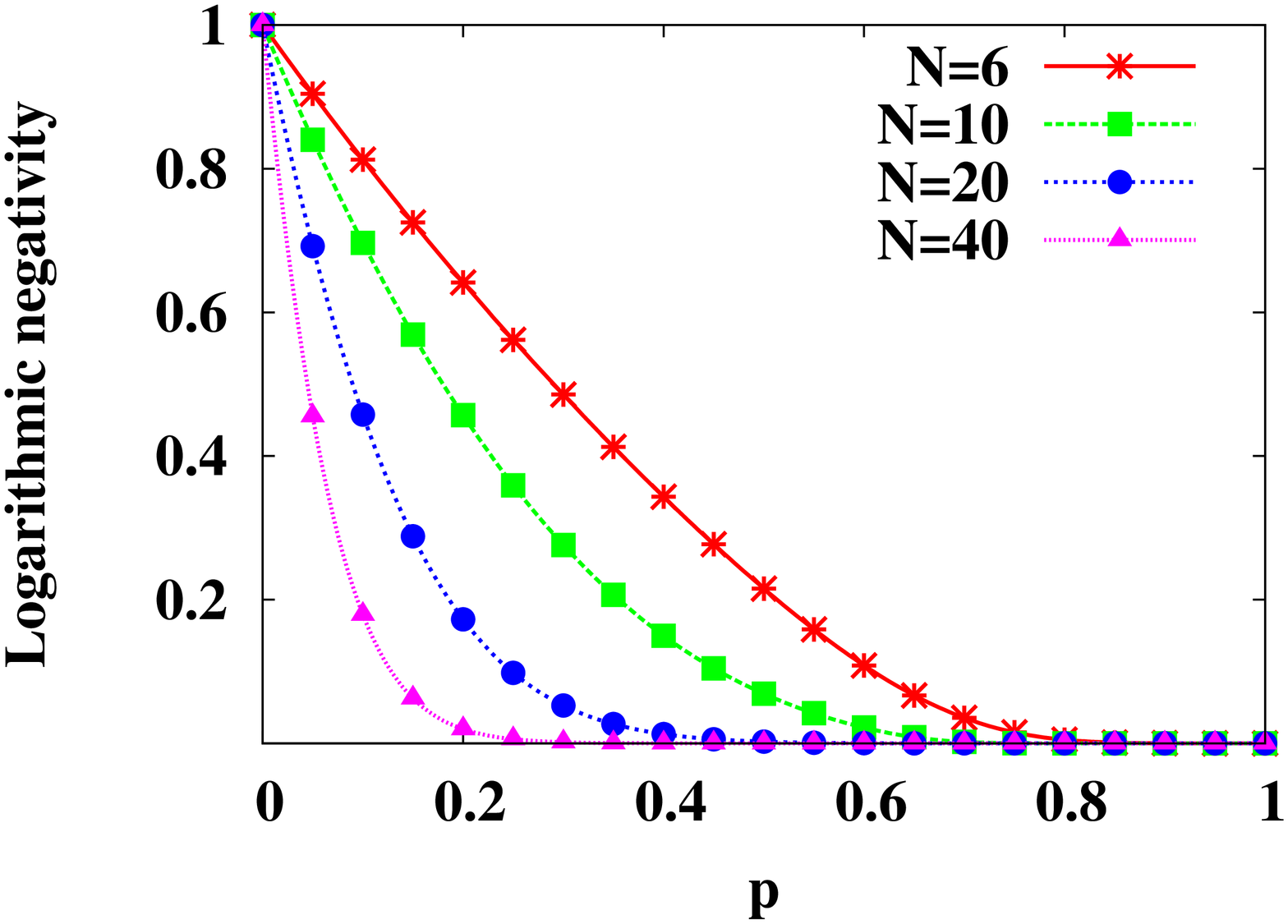}
\caption{(Color online) Entanglement of the noisy \(G\) state depends on the number of particles in the macroscopic sector. 
The logarithmic negativity (in ebits) of the noisy \(G\) state is plotted on the vertical axes against 
the decohering parameter \(p\) (dimensionless) on the horizontal axes, for different values of \(N\). 
The left panel is for the 
local phase
damping channels while the right one is for the local amplitude damping channels. We choose \(k=1\).}
\label{fig:ln-gs-pdc-adc-diff-n}
\end{figure}

 Let us begin by considering the effect of local phase damping channels on the state \(\arrowvert G \rangle _{\mu A_{1}...A_{N}}\). 
 In this case, the block which gives the negative eigenvalues is given by ($N$ is assumed to be even)
 \[B_{G}^{lpdc}= \frac{1}{2}\left( \begin{array}{cc}
0_{\frac{N}{2}\times \frac{N}{2}} & B_{\frac{N}{2}\times\frac{N}{2}} \\
B_{\frac{N}{2}\times\frac{N}{2}}&0_{\frac{N}{2}\times\frac{N}{2}} 
\end{array} \right),\]
 where
 \[B_{\frac{N}{2}\times\frac{N}{2}}= \left( \begin{array}{cccccc}
\gamma^{N-1} & \gamma^{N-1}&.&.&\gamma^{N-1}.&\gamma^{N+1} \\
\gamma^{N-1}&\gamma^{N-1}&.&.&\gamma^{N+1}&\gamma^{N-1} \\
.&.&.&\gamma^{N+1}&.&.\\
.&.&.&.&.&.\\
.&.&.&.&.&.\\
\gamma^{N+1}&\gamma^{N-1}&.&.&\gamma^{N-1}&\gamma^{N-1}
\end{array} \right),\]
and \(0_{\frac{N}{2}\times\frac{N}{2}}\) is an $\frac{N}{2}\times\frac{N}{2}$ matrix with all entries being \(0\). The negative eigenvalues are
\begin{equation*}
 \frac{1}{2N}\gamma^{N-1}(\gamma^{2}-1) \mbox{ with multiplicity } (N-1), 
\end{equation*}
\begin{equation*}
 \frac{1}{2N}(\gamma^{N-1}-\gamma^{N+1}) \mbox{ with multiplicity }(N-1),
\end{equation*}
and
\begin{equation*}
 \frac{1}{2N}(-(N-1)\gamma^{N-1}-\gamma^{N+1}) \mbox{ with multiplicity }  1.
\end{equation*}
From the eigenvalues, it is clear that the entanglement of the \(G\) state in the micro : macro bipartition, after it is affected by the local phase damping channels, depends on 
the size of macroscopic sector (see Fig. \ref{fig:ln-gs-pdc-adc-diff-n}). For \(N=6\), the \(G\) state can sustain  \(81\%\) local phase damping noise, 
which is lower than that of the  \(H_{C_{N}}^{m}\) for any \(N\) and for \(m=1,2,3\) (see Table I).

\begin{figure}
 \includegraphics[angle=270,width=4.2cm,totalheight=3.7cm]{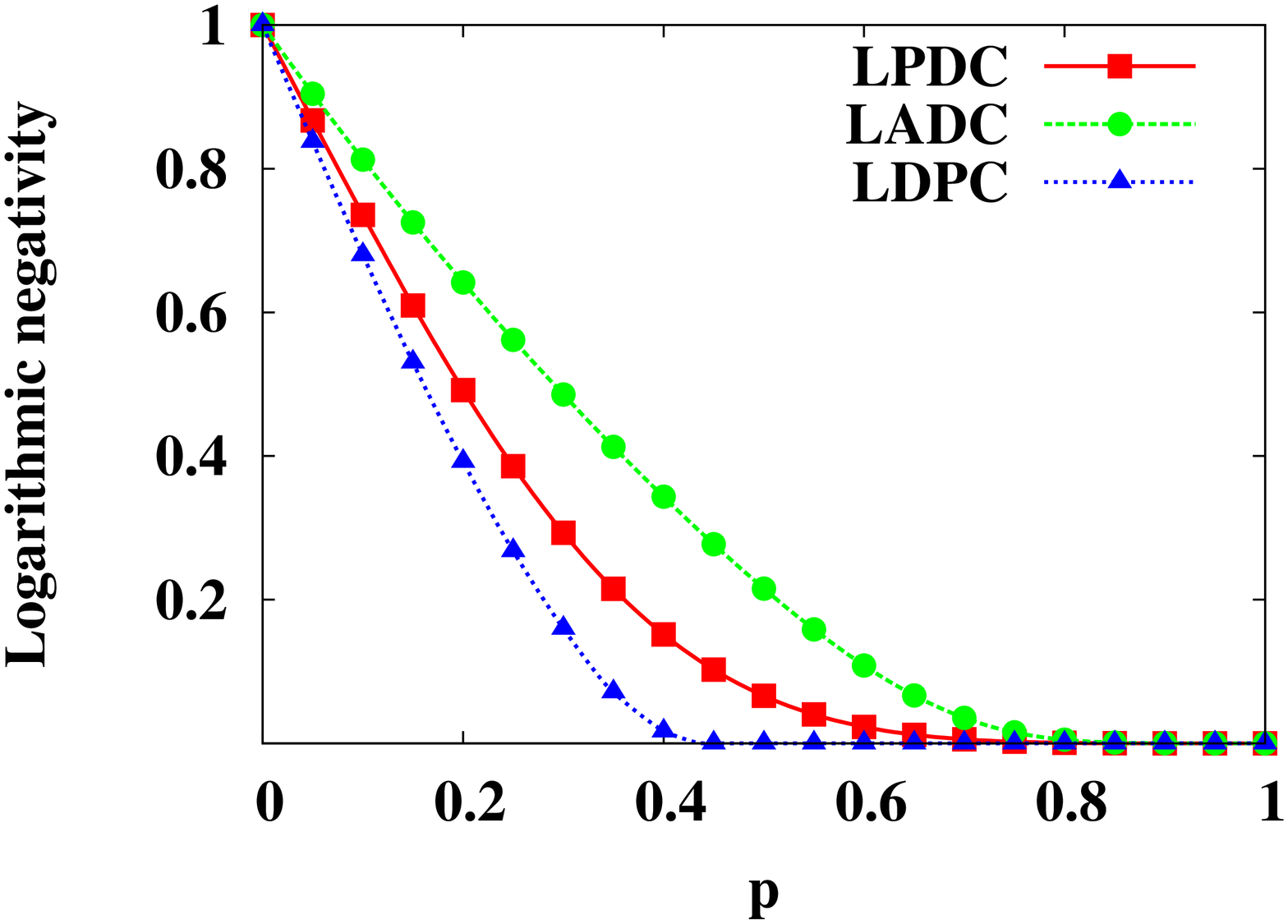}
  \includegraphics[angle=270,width=4.2cm,totalheight=3.7cm]{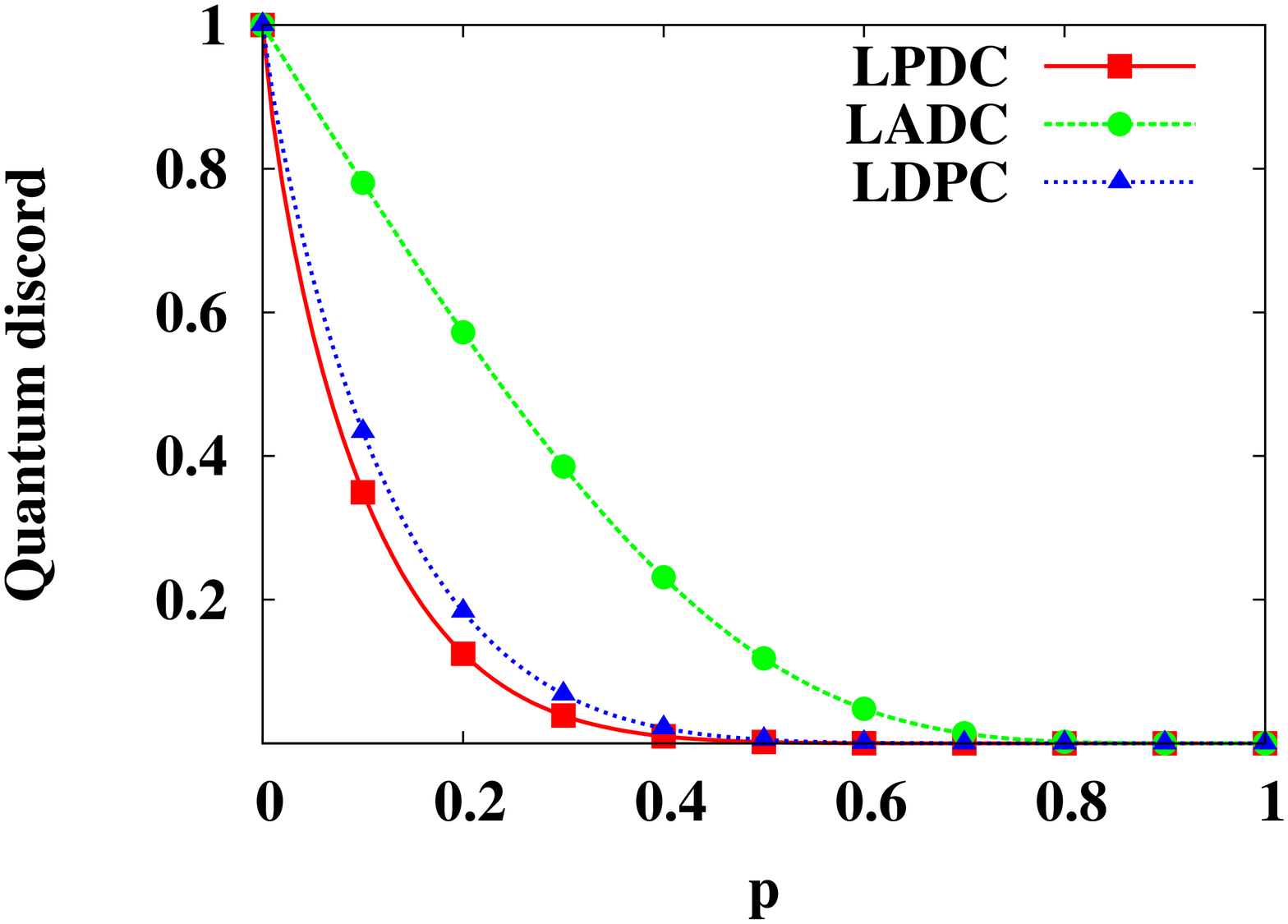}
 \caption{(Color online) The panels here are the same as in Fig. \ref{fig:ln-pdc-adc-dpc-hcs}, except that the curves here pertain to the noisy \(G\) states. 
 For numerical values, see Table I.
 }
  \label{fig:ln-qd-gs-pdc-adc-dpc-n6}
 \end{figure}

\begin{figure}
\includegraphics[angle=270,width=8.2cm,totalheight=4.5cm]{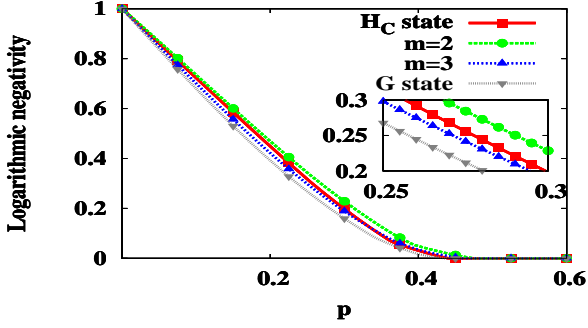}
\caption{(Color online) The plot here is the same as the left panel in Fig. \ref{fig:ln-dpc-ghz-HCS}, except that the curve for the 
GHZ state is replaced by that for the \(G\) state. The inset shows a magnified view of the curves in an intermediate zone.
 The numerical values of \(p\) at which the entanglements vanish, for the different states, are given in Table I.}
\label{fig:ghz-hcs-g-inset}
\end{figure}

We now consider the effect of local amplitude damping channels on the state. The block which gives the negative eigenvalues,
after partially transposing the noisy state, is of dimension \(2N\times2N\), and is given by
 \[B_{G}^{ladc}= \frac{1}{2}\left( \begin{array}{ccccccccccc}
l_{1}&.&.&.&l_{1}&l_{2}&.&.&.&l_{2} \\
.&.&.&.&.&.&.&.&.&.\\
.&.&.&.&.&.&.&.&.&.\\
.&.&.&.&.&.&.&.&.&.\\
l_{1}&.&.&.&l_{1}&l_{2}&.&.&.&l_{2} \\
l_{2}&.&.&.&l_{2}&l_{3}&l_{4}&.&.&l_{4} \\
.&.&.&.&.&.&.&.&.&.\\
.&.&.&.&.&.&.&.&.&.\\
.&.&.&.&.&.&.&.&.&.\\
l_{2}&.&.&.&l_{2}&l_{4}&l_{4}&.&.&l_{3}
\end{array} \right),\]
where
\(l_{1}=\frac{1}{N}p \gamma^{N-1}\), \(l_{2}=\frac{1}{N}\gamma^{\frac{N+1}{2}}\), \(l_{3}=\frac{(N-1)}{N}\gamma^{2}p^{N-2}\), and \(l_{4}=\frac{1}{N}\gamma^{2}p^{N-2}\).
The negative eigenvalues are
\begin{align}
 \lambda_{G,1}^{ladc}&=\frac{1}{4}[(Nl_{1}+l_{3}+(N-1)l_{4})\nonumber \\
 &-\sqrt{4N^{2}l_{2}^{2}+(-Nl_{1}+l_{3}+(N-1)l_{4})^{2}}],
\end{align}
and 
\begin{align}
 \lambda_{G,2}^{ladc}&=\frac{1}{4}[a_{1}+c_{1}+(2N-4)d_{1}\nonumber\\
 &-\sqrt{2N(N-1)b_{1}^{2}+(-a_{1}+c_{1}+(2N-4)d_{1})^{2}}].
\end{align}
where
\(a_{1}=p^{N-1}\gamma\), \(b_{1}=\frac{2}{N}\gamma^{\frac{N-1}{2}}p\), \(c_{1}=\frac{2}{N} p^{2}\gamma^{N-2}\), and \(d_{1}=\frac{1}{N}p^{2}\gamma^{N-2}\).
Fig. \ref{fig:ln-gs-pdc-adc-diff-n} clearly shows that the entanglement, as quantified by the  logarithmic negativity, of the \(G\) state decreases with the 
increase in the number of particles in the macroscopic section, unlike in the case of the \(H_{C_{N}}^{m}\) state, for both local phase and local amplitude damping channels.
A comparison of the logarithmic negativities and quantum discords  for the \(G\) state under different local noisy channels is presented in Fig. \ref{fig:ln-qd-gs-pdc-adc-dpc-n6} where we additionally consider the 
local depolarizing channel.
For a fixed \(N\), from the perspective of the robustness of entanglement, the local amplitude damping channel is the best among the channels considered, while the local depolarizing channel is the worst.
The situation is rather similar for quantum discord -- local amplitude damping is still the best, but the local phase damping is marginally 
worse than local depolarizing. Note the qualitative similarity of this 
situation with that for the \(H_{C_{N}}^{m}\) state (see Fig. \ref{fig:ln-pdc-adc-dpc-hcs}) and the dissimilarity with that for the GHZ state (see Fig. \ref{fig:ln-adc-pdc-dpc-ghz}). 
In Fig.  \ref{fig:ghz-hcs-g-inset}, we compare the entanglement of the local depolarized \(G\) state with those in the 
local depolarized \(H_{C_{N}}^{m}\) states.
We find that the critical value for the \(G\) state is rather similar to those of the \(H_{C_{N}}^{m}\) states.
See Table I for further details. 
Note that the plots of logarithmic negativity and quantum discord of the \(G\) state are shown here for \(k=1,N=6\) under the different noisy channels. The corresponding analytic expressions
of logarithmic negativity,
for local phase damping and local amplitude damping, for arbitrary \(N\) and \(k\), are given in the text. For the case  of the local depolarizing channel, the logarithmic negativity is calculated
numerically. The optimization for quantum discord is performed numerically in all the cases.

\subsection{\(H_{C_{N}}^{N-1}\) state}
Let us introduce another multiparticle state which is quite similar to the \(H_{C_{N}}^1\) state, with only the 
$ W_{N}^1$ replaced by $ W_{N}^{N-1}$, where $ W_{N}^{N-1}$ is obtained from Eq. (\ref{eq:wnm}) by putting \(m=N-1\). 
The state, therefore,  is given by
\begin{align}
\label{eq:state-2}
 \arrowvert H_{C_{N}}^{N-1}\rangle _{\mu A_{1}...A_{N}}=
&\frac{1}{\sqrt{2}}(\arrowvert 0 \rangle _{\mu} \arrowvert W_{N}^{N-1} \rangle _{A_{1}\ldots A_{N}} \nonumber\\
&+\arrowvert 1 \rangle _{\mu} \arrowvert 0 \ldots 0 \rangle_{A_{1}\ldots A_{N}}).
\end{align}
This state is a cat-like state in the same sense as the \(H_{C_{N}}^1\). 
Moreover, the states \( \arrowvert W_{N}^{N-1} \rangle\) and 
\(\arrowvert 0 \ldots 0 \rangle\) are macroscopically different in terms of their \(\sigma_z\)-magnetizations.

\begin{figure}
 \includegraphics[angle=270,width=4.2cm,totalheight=3.5cm]{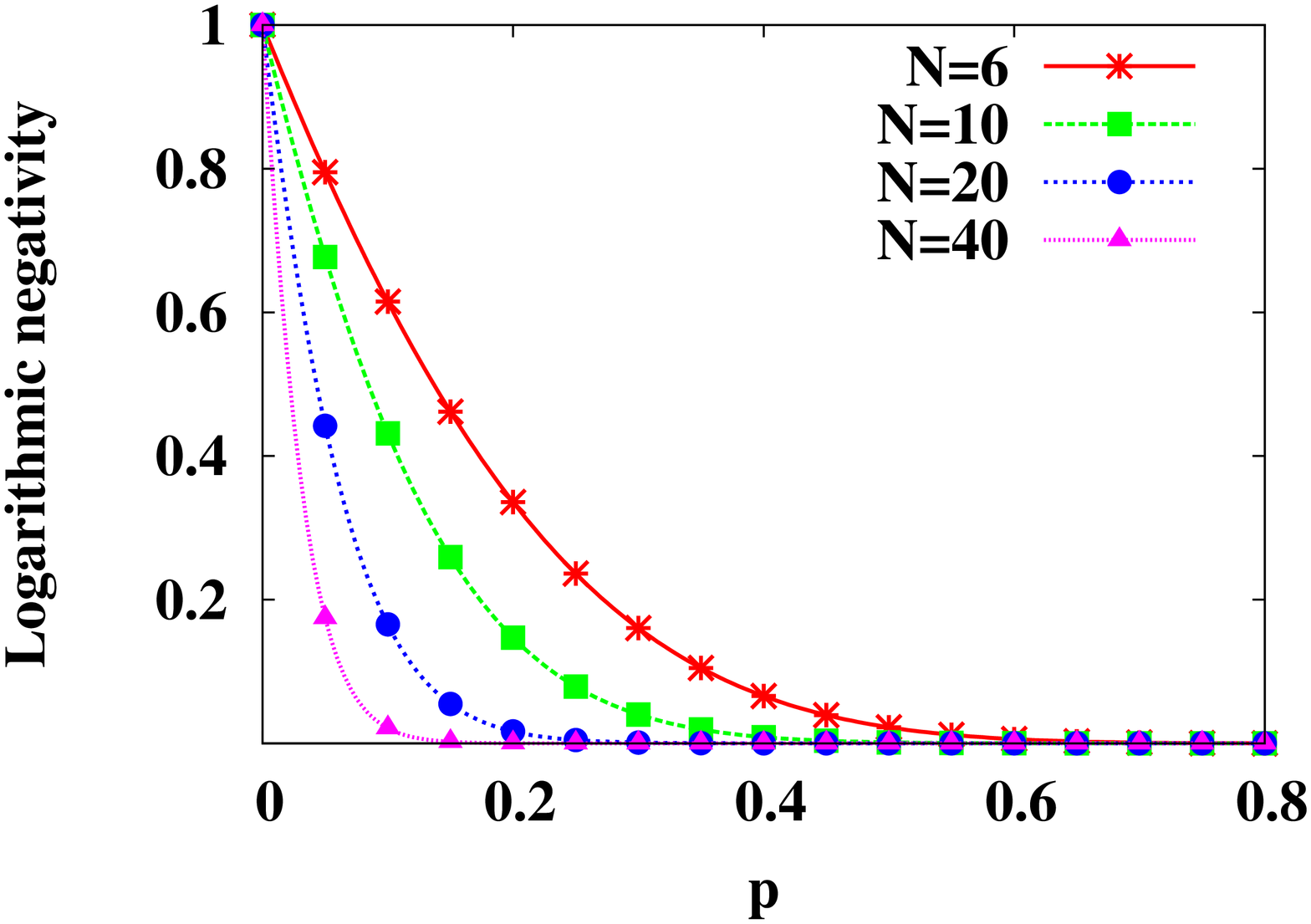}
  \includegraphics[angle=270,width=4.2cm,totalheight=3.5cm]{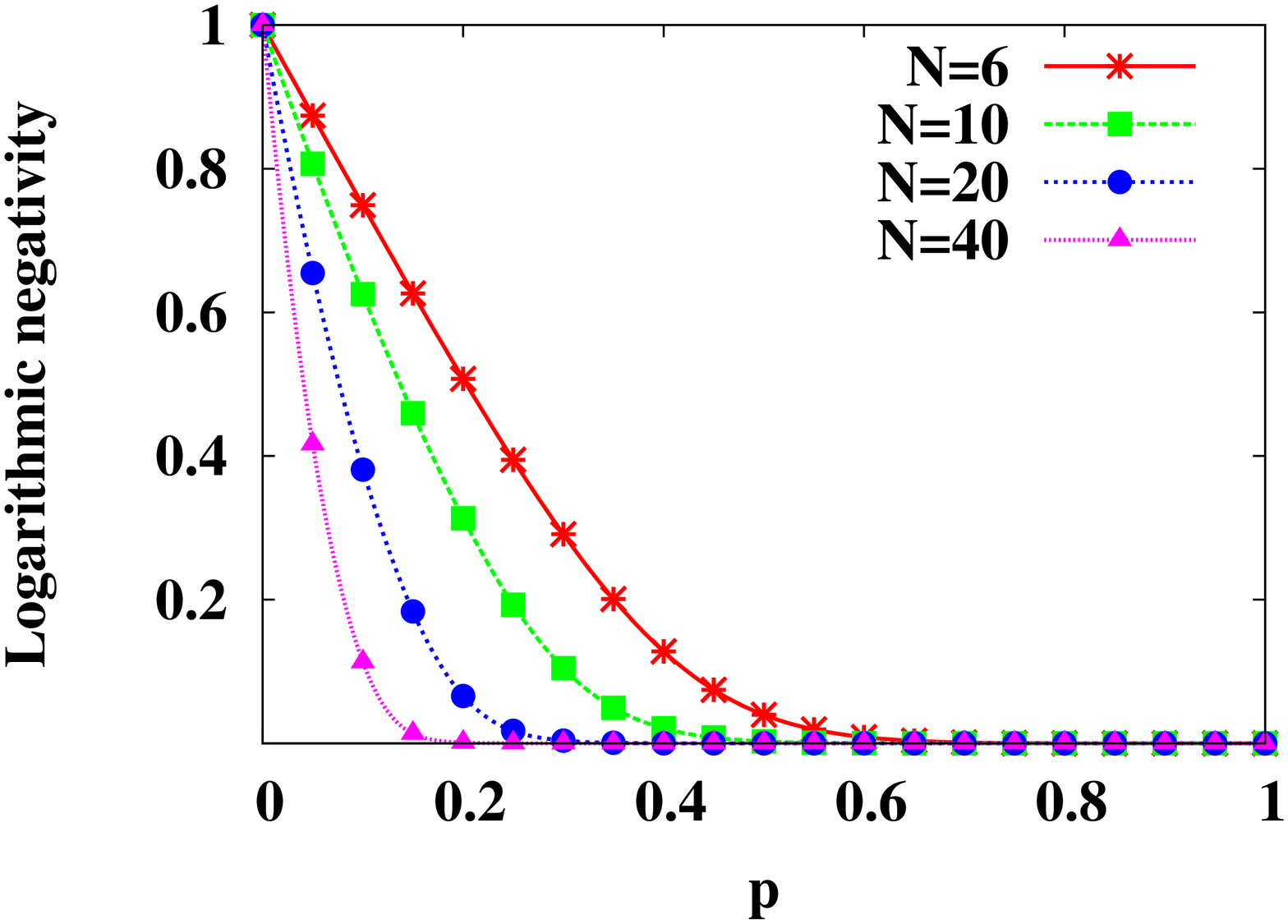}
 \caption{(Color online) The panels in this figure are the same as in Fig. \ref{fig:ln-gs-pdc-adc-diff-n}, except that here they are for the 
\(H_{C_{N}}^{N-1}\) state.
}
 \label{fig:qwerty}
  \end{figure}

The effects of local phase  and amplitude damping channels, for this case can be obtained by putting $m=N-1$ in Eqs. 
(\ref{eq:lpdcnm}) and (\ref{eq:ln-eq-adc-hcs}). (See Fig. \ref{fig:qwerty}.)
Let us now investigate the effect of the local depolarizing channels on the state. The blocks which give the negative eigenvalues are of dimension \((N+1)\times(N+1)\), and are 
  \begin{figure}
 \includegraphics[angle=270,width=4.2cm,totalheight=3.5cm]{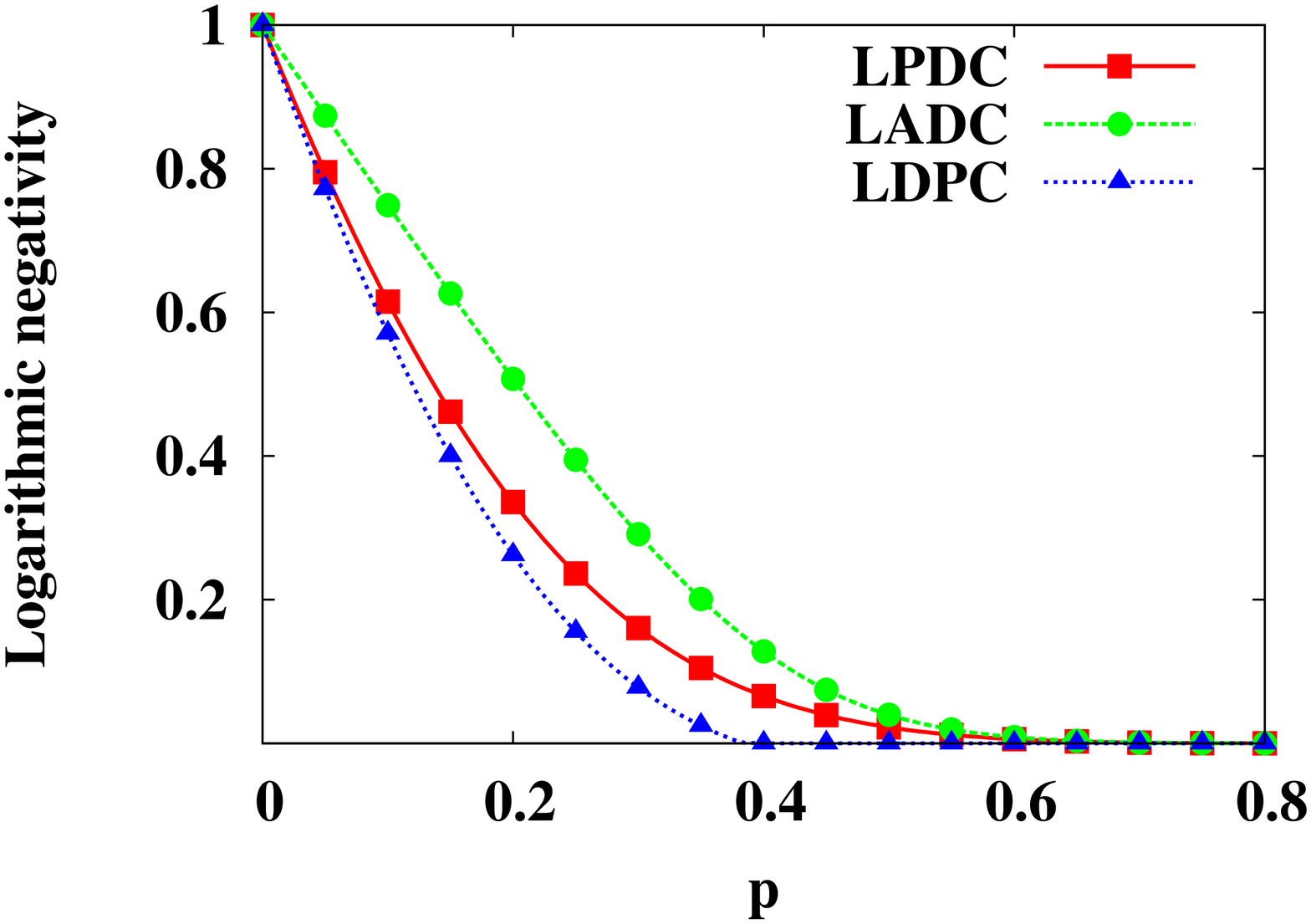}
  \includegraphics[angle=270,width=4.2cm,totalheight=3.5cm]{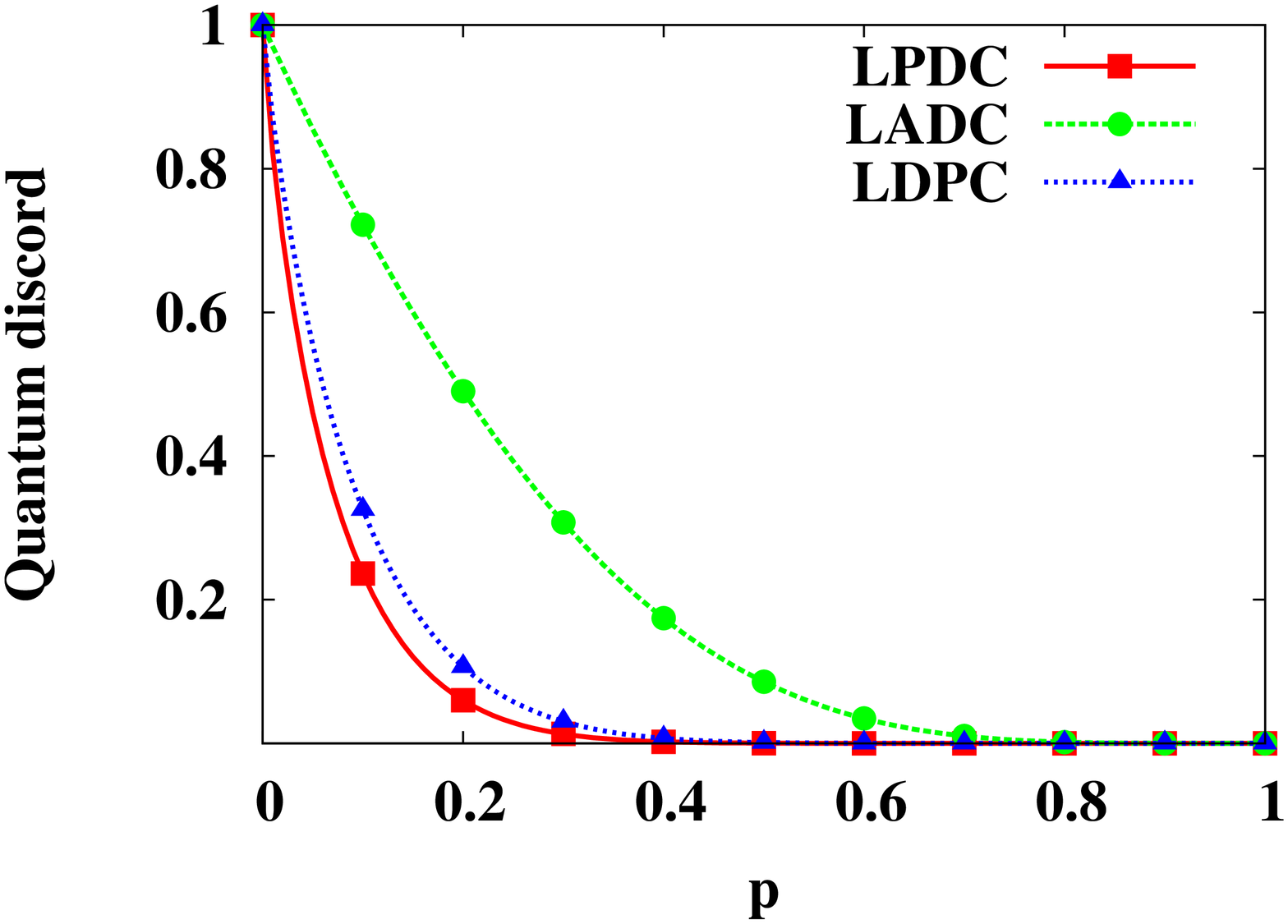}
\caption{(Color online) The panels in this figure are the same as in Fig. \ref{fig:ln-pdc-adc-dpc-hcs}, except that here they pertain to the 
\(H_{C_{N}}^{N-1}\) state, for \(N=6\) and \(k=1\). 
}
 \label{fig:ln-dpc-adc-pdc-s2}
 \end{figure}
 \[B_{H_{C_{N}}^{N-1}}^{1}= \frac{1}{2}\left( \begin{array}{ccccccc}
a & b&b&.&.&.&b \\
b&c&d&.&.&.&.d \\
b&d&c&.&.&.&d\\
.&.&.&.&.&.&d\\
.&.&.&.&.&.&.\\
.&.&.&.&.&.&.\\
.&.&.&.&.&.&.\\
b&d&d&.&.&.&c
\end{array} \right)\]
and 
\[B_{H_{C_{N}}^{N-1}}^{2}= \frac{1}{2}\left( \begin{array}{ccccccc}
\tilde{a} &\tilde{b}&\tilde{b}&.&.&.&\tilde{b} \\
\tilde{b}&\tilde{c}&\tilde{d}&.&.&.&.\tilde{d} \\
\tilde{b}&\tilde{d}&\tilde{c}&.&.&.&\tilde{d}\\
.&.&.&.&.&.&\tilde{d}\\
.&.&.&.&.&.&.\\
.&.&.&.&.&.&.\\
.&.&.&.&.&.&.\\
\tilde{b}&\tilde{d}&\tilde{d}&.&.&.&\tilde{c}
\end{array} \right),\]
respectively,
where \(a=\beta \alpha^{N}+\frac{1}{N}\alpha^{2}\beta^{N-1}\), \(b=\frac{1}{\sqrt{N}}\gamma^{N}\alpha\), 
\(c=\alpha^{2}\beta^{N-1}+\frac{1}{N}(\beta\alpha^{N}+(N-1)\beta^{3}\alpha^{N-2})\), \(d=\frac{1}{N}\beta \gamma^{2}\alpha^{N-2}\), and
\(\tilde{a}=\alpha \beta^{N}+\beta^{2} \alpha^{N-1}\), 
\(\tilde{b}= \frac{1}{\sqrt{N}}\beta\gamma^{N}\), \(\tilde{c}= \beta^{2}\alpha^{N-1}+\frac{1}{N}(\alpha\beta^{N}+(N-1)\alpha^{3}\beta^{N-2})\), \(\tilde{d}=\frac{1}{N}\beta\gamma^{2}\alpha^{N-2}\).
The negative eigenvalues are 
\begin{align}
 \lambda_{H_{C_{N}}^{N-1},1}^{ldpc}&=\frac{1}{2}(a+c+(N-1)d\nonumber\\
 &-\sqrt{(4Nb^{2}+(-a+c+(N-1)d)^{2})})
\end{align}
and 
\begin{align}
 \lambda_{H_{C_{N}}^{N-1},2}^{ldpc}&=\frac{1}{4}(\tilde{a}+\tilde{c}+(N-1)\tilde{d}\nonumber\\
 &-\sqrt{(4N\tilde{b}^{2}+(-\tilde{a}+\tilde{c}+(N-1)\tilde{d})^{2})})
\end{align}
respectively.
The logarithmic negativity of the local depolarized state is therefore given by
\begin{align}
 E_{H_{C_{N}}^{N-1}}^{ldpc}&=\mbox{log}_{2}[2|(\mbox{min}(0,\lambda_{H_{C_{N}}^{N-1},1}^{ldpc})| \nonumber\\
 +& 2|\mbox{min}(0,\lambda_{H_{C_{N}}^{N-1},2}^{ldpc})|+1].
\end{align}
The comparison among local amplitude damping, local phase damping, and local depolarizing channels, for the case when \(k=1\) and \(N=6\), i.e.
for the \(H_{C_{6}}^{5}\) state, is presented in 
Fig. \ref{fig:ln-dpc-adc-pdc-s2}. It is clear from the  figure
that the effect of the local amplitude damping channel, on the \(H_{C_{N}}^{N-1}\) state is much less pronounced as 
compared to local phase damping and local depolarizing channels. For fixed noise models, the percentage of noise that the individual states can sustain, before it becomes separable in the microscopic to macroscopic partition, is given in Table I.
 
\section{conclusion}
\label{conclusion}
Studying quantum systems under environmental noise is important from a 
variety of perspectives ranging from fundamental concepts, like the 
quantum to classical transition, to robustness of quantum information 
processing and computational tasks. We have studied a class of 
multipartite quantum states, which are quantum superpositions in a 
composite system, consisting of a microscopic and a macroscopic part. The 
microscopic part is assumed to be formed by a few qubits, while the 
macroscopic one is built by a large number of the same. We have investigated the 
effect of several paradigmatic models of \emph{local} environmental noise on 
the multiparty states, by calculating quantum correlations between their 
microscopic and macroscopic sectors, after the states are affected by the 
local noise. We have considered three different types of local noisy 
channels, viz. the local phase damping, the local amplitude damping, and 
the local depolarizing channels. In studying the quantum correlations, we 
consider both entanglement measures as well as information theoretic 
quantum correlation measures.

We find that the quantum correlations of all the states from the class 
considered here remain independent of the size of the macroscopic sector under 
local phase damping and local amplitude damping channels. We identify the 
state in this class which remains maximally robust for a given 
local noise. Interestingly, we observe that for all the quantum states in 
the class, entanglement is almost equally robust against local 
amplitude damping and local phase damping noise, while being much worse 
off against local depolarization. In contrast, quantum discord is much more 
robust against local amplitude damping than local depolarization or local 
phase damping noise. Finally, we find that the quantum correlations in the proposed class of multiparty 
quantum states is better preserved than that in the other macroscopic superposition states 
against all the local noise models. The findings may 
help us to identify a potential candidate for quantum memory
devices.

\end{document}